\documentclass[aps,pra,twocolumn,superscriptaddress]{revtex4-1}

\usepackage[utf8]{inputenc}  
\usepackage[T1]{fontenc}    
\usepackage{hyperref}       
\usepackage{url}          
\usepackage{booktabs}        
\usepackage{amsfonts}        
\usepackage{nicefrac}        
\usepackage{microtype}       
\usepackage{bm}
\usepackage{braket}
\usepackage{graphicx}
\usepackage{amsthm}
\usepackage{amssymb}
\usepackage{amsmath}

\usepackage{algorithm}
\usepackage{algorithmic}

\usepackage{appendix}

\usepackage{adjustbox}
\usepackage[normalem]{ulem}
\newtheorem{thm}{Theorem}
\newtheorem{lem}{Lemma}

\usepackage{mathtools}
\usepackage{bbm}
\usepackage[explicit]{titlesec}
\usepackage{nccmath}
\usepackage{xcolor}
\usepackage{caption}
 
\usepackage{subcaption}
\usepackage{wrapfig}
\DeclareMathOperator{\Tr}{Tr}
\DeclareMathOperator{\RX}{R_X}
\DeclareMathOperator{\RY}{R_Y}
\DeclareMathOperator{\RZ}{R_Z}

\DeclareMathOperator{\CNOT}{CNOT}

\DeclareMathOperator{\Ha}{\text{Ha}}

 \usepackage{color}
 
 \definecolor{codegreen}{rgb}{0,0.6,0}
\definecolor{codegray}{rgb}{0.5,0.5,0.5}
\definecolor{codepurple}{rgb}{0.58,0,0.82}
\definecolor{backcolour}{rgb}{0.95,0.95,0.92}
 \usepackage{listings}
 \lstdefinestyle{mystyle}{
    backgroundcolor=\color{backcolour},   
    commentstyle=\color{codegreen},
    keywordstyle=\color{magenta},
    numberstyle=\tiny\color{codegray},
    stringstyle=\color{codepurple},
    basicstyle=\ttfamily\footnotesize,
    breakatwhitespace=false,         
    breaklines=true,                 
    captionpos=b,                    
    keepspaces=true,                 
    numbers=left,                    
    numbersep=5pt,                  
    showspaces=false,                
    showstringspaces=false,
    showtabs=false,                  
    tabsize=2
}
 
\begin{document}

\title{Quantum circuit architecture search for variational quantum algorithms}

\author{Yuxuan Du}
\email{duyuxuan123@gmail.com}
 \affiliation{UBTECH Sydney AI Centre, School of Computer Science, Faculty of Engineering, The University of Sydney, NSW 2008, Australia}
 \affiliation{JD Explore Academy, Beijing 101111, China}

\author{Tao Huang}
\affiliation{SenseTime, Beijing 100080, China}

\author{Shan You}
\affiliation{SenseTime, Beijing 100080, China}

\author{Min-Hsiu Hsieh}
\email{min-hsiu.hsieh@foxconn.com} 
\affiliation{Centre for Quantum Software and Information, Faculty of Engineering and Information Technology, University of Technology Sydney, NSW 2007, Australia}
\affiliation{Hon Hai Quantum Computing Research Center, Taipei 114, Taiwan}

\author{Dacheng Tao}
 \email{dacheng.tao@sydney.edu.au}
\affiliation{UBTECH Sydney AI Centre, School of Computer Science, Faculty of Engineering, The University of Sydney, NSW 2008, Australia}
 \affiliation{JD Explore Academy, Beijing 101111, China}

\date{\today}

\maketitle

\textbf{Variational quantum algorithms (VQAs) are expected to be a path to quantum advantages on noisy intermediate-scale quantum devices. However, both empirical and theoretical results exhibit that the deployed ansatz heavily affects the performance of VQAs such that an ansatz with a larger number of quantum gates enables a stronger expressivity, while the accumulated noise may render a poor trainability. To maximally improve the robustness and trainability of VQAs, here we devise a resource and runtime efficient scheme termed quantum architecture search  (QAS). In particular, given a learning task, QAS automatically seeks a near-optimal ansatz (i.e., circuit architecture) to balance benefits and side-effects brought by adding more noisy quantum gates  to achieve a good performance.  We implement QAS on both the numerical simulator and real quantum hardware, via the IBM cloud, to accomplish data classification and quantum chemistry tasks. In the problems studied, numerical and experimental results show that QAS can not only alleviate the influence of quantum noise and barren plateaus, but also outperforms VQAs with pre-selected ansatze.  }

The variational quantum learning algorithms (VQAs) \cite{cerezo2020variational2,bharti2021noisy}, including quantum neural network \cite{beer2020training,farhi2018classification,schuld2019quantum} and variational quantum eigen-solvers \cite{peruzzo2014variational,wang2019accelerated,stokes2020quantum,mitarai2019generalization}, are a class of promising candidates to use noisy intermediate-scale quantum (NISQ) devices to solve practical tasks that are  beyond the reach of classical computers \cite{preskill2018quantum}. Recently, the effectiveness of VQAs towards small-scale learning problems such as low-dimensional synthetic data classification, image generation, and energy estimation for small molecules has been validated by experimental studies \cite{havlivcek2019supervised,huang2021experimental,kandala2017hardware,google2020hartree}. Despite the promising achievements, the performance of VQAs will degrade significantly when the qubit number and circuit depth become large, caused by the trade-off between the expressivity and trainability \cite{holmes2021connecting}. More precisely, under the NISQ setting, involving more quantum resources (e.g., quantum gates) to implement the ansatz results in both a  positive and negative aftermath. On the one hand, the expressivity of the ansatz, which determines whether the target concept will be covered by the represented hypothesis space, will be strengthened by increasing the number of trainable gates    \cite{benedetti2019parameterized,caro2021generalization,du2018expressive,du2021efficient}. On the other hand, a deep circuit depth implies that the gradient information received by the classical optimizer is full of noise and the valid information is exponentially vanished, which may lead to divergent optimization or barren plateaus  \cite{du2020learnability,cerezo2020cost,mcclean2018barren,sweke2019stochastic,wang2020noise}. With this regard, it is of great importance to design an efficient approach to dynamically control the  expressivity and trainability of VQAs to attain good performance.

Initial studies have developed to two leading strategies to address the above issue. The first one is quantum error mitigation techniques. Representative methods to suppress the noise effect on NISQ machines are  quasi-probability    \cite{temme2017error,endo2018practical}, extrapolation \cite{li2017efficient}, quantum subspace expansion \cite{mcclean2017hybrid}, and data-driven methods \cite{strikis2020learning,czarnik2020error}. In parallel to quantum error mitigation, another way is constructing ansatz with a variable structure. Compared with traditional VQAs with the fixed ansatz, this approach can not only maintain a shallow depth to suppress noise and trainability issues, but also keep sufficient expressibility to contain the solution. Current literature generally adopts brute-force strategies to design such a variable ansatz \cite{chivilikhin2020mog,li2020quantum,ostaszewski2021structure}. This implies that the required computational overhead is considerable, since the candidates of possible ansatze scale exponentially with respect to the qubits count and the circuit depth. How to efficiently seek a near-optimal ansatz remains largely unknown.

\begin{figure*}[ht!] 
\centering
\includegraphics[width=0.96\textwidth]{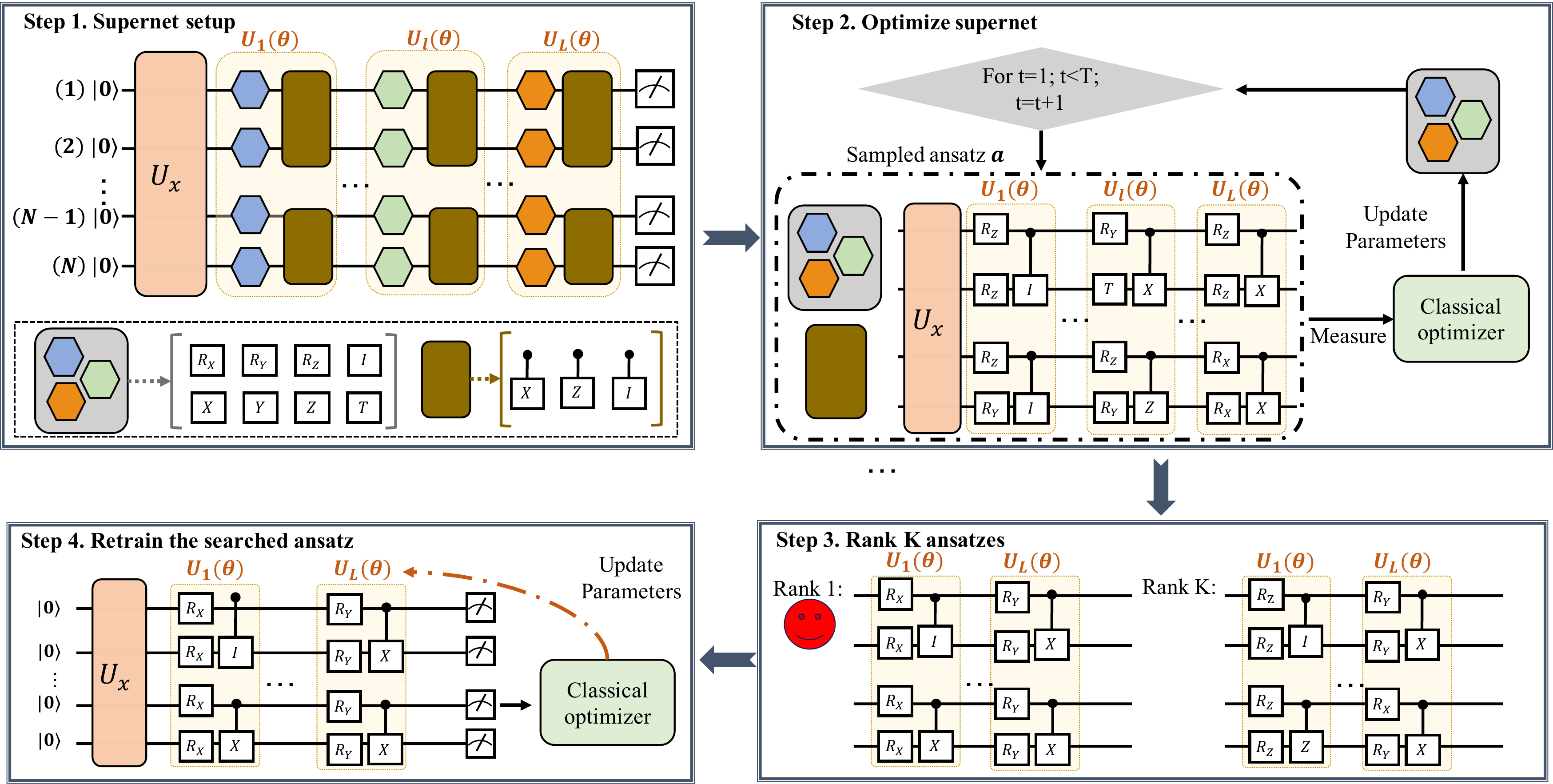}
\caption{\small{\textbf{Paradigm of the quantum architecture search scheme (QAS).} In Step 1, QAS sets up supernet $\mathcal{A}$, which defines the ansatze pool $\mathcal{S}$ to be searched and parameterizes each ansatz in $\mathcal{S}$ via the specified weight sharing strategy. All possible single-qubit gates are highlighted by hexagons and two-qubit gates are highlighted by the brown rectangle. The unitary $U_{\bm{x}}$ refers to the data encoding layer. In Step 2, QAS optimizes the trainable parameters for all candidate ansatzes. Given the specified learning task $\mathcal{L}$, QAS iteratively samples an ansatz $\bm{a}^{(t)}\in\mathcal{S}$ from $\mathcal{A}$ and optimizes its trainable parameters to minimize $\mathcal{L}$. $\mathcal{A}$ correlates parameters among different  ansatzes via weight sharing strategy. After $T$ iterations, QAS moves to Step 3 and exploits the trained parameters $\bm{\theta}^{(T)}$ and the predefined $\mathcal{L}$ to compare the performance among $K$ ansatze. The ansatz with the best performance is selected as the output, indicated by a red smiley face. Last, in Step 4, QAS utilizes the searched ansatz and the  parameters $\bm{\theta}^{(T)}$ to retrain the quantum solver with few  iterations.   }              }   
\label{fig:QAS}
\end{figure*} 

In this study, we devise a quantum architecture search scheme (QAS) to effectively generate variable structure ansatze, which considerably improves  the learning performance of VQAs. The advantage of QAS is ensured by unifying the noise inhibition and the enhancement of trainability for VQAs as a learning problem. In doing so,  QAS does not request any ancillary quantum resource and its runtime is almost the same as conventional VQA-based algorithms. Moreover, QAS is compatible with all quantum platforms, e.g., optical, trapped-ion, and superconducting quantum machines, since it can actively adapt to physical restrictions and weighted noise of varied quantum gates. In addition, QAS can seamlessly integrate with other quantum error mitigation methods \cite{li2017efficient,endo2018practical,temme2017error} and solutions of resolving barren plateaus   \cite{cerezo2020cost,grant2019initialization,skolik2021layerwise,zhang2021toward}. Celebrated by the universality and efficacy,  QAS contributes to a broad class of VQAs on various quantum machines.  

\medskip
\medskip 
\noindent \textbf{RESULTS}\\
\noindent\textbf{The mechanism of VQAs.} Before moving on to present QAS, we first recap the mechanism of VQAs. Given an input $\mathcal{Z}$ and an objective function $\mathcal{L}$, VQA employs a gradient-based classical optimizer that continuously updates parameters in an ansatz (i.e., a parameterized quantum circuit) $U(\bm{\theta})$ to find the optimal $\bm{\theta}^*$, i.e.,  
\begin{equation}\label{eqn:exp_erm}
\bm{\theta}^*= \arg \min_{\bm{\theta}\in\mathcal{C}} \mathcal{L}(\bm{\theta},\mathcal{Z}),
\end{equation} 
where $\mathcal{C}\subseteq \mathbb{R}^d$ is a constraint set,  and $\bm{\theta}$ are adjustable parameters of quantum gates \cite{benedetti2019parameterized,du2018expressive}. For instance, when VQA is specified as an Eigen-solver \cite{peruzzo2014variational},  $\mathcal{Z}$ refers to a Hamiltonian and the objection function could be chosen as $\mathcal{L}=\Tr(\mathcal{Z}\ket{\psi(\bm{\theta})}\bra{\psi(\bm{\theta})})$, where $\ket{\psi(\bm{\theta})}$ is the quantum state generated by $U(\bm{\theta})$. For compatibility, throughout the whole study, we focus on exploring how QAS enhances the trainability of one typical heuristic ansatz---hardware-efficient ansatz \cite{havlivcek2019supervised,kandala2017hardware}. Such an ansatz is supposed to obey a multi-layer layout,
\begin{equation}\label{eqn:UL_def}
	U(\bm{\theta})=\prod_{l=1}^LU_l(\bm{\theta})\in  SU(2^N),
\end{equation}  
where $U_l(\bm{\theta})$ consists of a sequence of parameterized single-qubit and two-qubit quantum gates, and $L$ denotes the layer number. Note that the arrangement of quantum gates in $U_l(\bm{\theta})$ is flexible, enabling VQAs to adequately use available quantum resources and to accord with any physical  restriction. Remarkably, the achieved results can be effectively extended to other representative ansatze. 

\medskip
\noindent\textbf{The scheme of quantum  architecture search.} Let us formalize the noise inhibition and trainability enhancement for VQAs as a learning task. Denote the set $\mathcal{S}$ as the ansatze pool that contains all possible ansatze (i.e., circuit architectures) to build $U(\bm{\theta})$ in Eqn.~(\ref{eqn:UL_def}). The size of $\mathcal{S}$ is determined by the qubits count $N$, the maximum circuit depth $L$,  and  the number of allowed types of quantum gates $Q$, i.e., $|\mathcal{S}|=O(Q^{NL})$. Throughout the whole study, when no confusion occurs, we denote $\bm{a}$ as the $a$-th ansatz $U(\bm{\theta},\bm{a})$ in $\mathcal{S}$. Notably, the performance of VQAs heavily relies on the employed ansatz selected from $\mathcal{S}$.  Suppose the quantum system noise, induced by $\bm{a}$, is modeled by the quantum channel $\mathcal{E}_{\bm{a}}$. Taking into account of the circuit architecture information and the related noise, the objective of VQAs can be  rewritten as 
\begin{equation}\label{eqn:obj_QAS}
	(\bm{\theta}^*,\bm{a}^*)= \arg \min_{\bm{\theta}\in\mathcal{C}, \bm{a}\in\mathcal{S}} \mathcal{L}(\bm{\theta}, \bm{a}, \mathcal{Z},\mathcal{E}_{\bm{a}}).
\end{equation} 
The learning problem formulated in Eqn.~(\ref{eqn:obj_QAS}) forces the optimizer to output the best quantum circuit architecture $\bm{a}^*$ by assessing both the effect of noise and the trainability. Notably,  Eqn.~(\ref{eqn:obj_QAS}) is  intractable via the two-stage optimization strategy that is broadly used in previous literature \cite{chivilikhin2020mog,li2020quantum,ostaszewski2021structure}, i.e., individually optimizing all possible ansatze from scratch and then ranking them to obtain $(\bm{\theta}^*,\bm{a}^*)$. This is because the classical optimizer needs to store and update $O(dQ^{NL})$ parameters, which forbids its applicability towards large-scale problems in terms of $N$ and $L$.

The proposed QAS belongs to the   one-stage optimization  strategy. Different from the two-state optimization strategy that suffers from the  computational bottleneck, this strategy ensures the efficiency of QAS. In particular, for the same number of iterations $T$, the memory cost of QAS is at most $T$ times more than that of conventional VQAs. Meanwhile, their runtime complexity is identical.    The protocol of QAS is shown in Figure~\ref{fig:QAS}. Two key elements of QAS are supernet and weight sharing strategy. Both of them contribute to locate a good estimation of $(\bm{\theta}^*,\bm{a}^*)$ within a reasonable runtime and memory usage.  Intuitively, weight sharing strategy in QAS refers to  correlating parameters among different ansatze. In this way, the parameter space, which amounts to the total number of trainable parameters required to be optimized in Eqn.~(\ref{eqn:obj_QAS}), can be effectively reduced. As for supernet, it  plays two significant roles in QAS: 1) supernet serves as the ansatz indicator, which defines the ansatze pool $\mathcal{S}$ (e.g., determined by the maximum circuit depth and the choices of quantum gates) to be searched; 2) supernet parameterizes each ansatz in $\mathcal{S}$ via the specified weight sharing strategy. QAS includes four steps, i.e., initialization (supernet setup), optimization, ranking, and fine tuning. We now elucidate these four steps. 

1. (Initialization.)    
 QAS employs a supernet $\mathcal{A}$ as an indicator for the ansatze pool $\mathcal{S}$. Concretely, the setup of the supernet $\mathcal{A}$ amounts to leveraging the indexing technique to track $\mathcal{S}$ using a linear memory cost. For instance, when $N=4$, $L=1$, and the choices of the quantum gates are $\{\RX,\RY,\RZ\}$ with $Q=3$, $\mathcal{A}$ indexes $\RX,\RY,\RZ$ as $`0$', `$1$', `$2$', respectively. With setting the range of $a,b,c,d$ as $\{0, 1, 2\}$,  the index list [`a', `b', `c', `d'] tracks $\mathcal{S}$, e.g., [`0', `0', `0', `0'] describes the ansatz $\otimes_{i=1}^4 \RX(\bm{\theta}_i)$  and [`2', `2', `2', `2'] describes the ansatz $\otimes_{i=1}^4 \RZ(\bm{\theta}_i)$. See Method for the construction of the ansatze pool $\mathcal{S}$ involving two-qubit gates. Meantime, as detailed below, $\mathcal{A}$ parameterizes all candidate ansatze via weight sharing strategy to reduce parameter space.

2. (Optimization.) QAS jointly optimizes $\{(\bm{a},\bm{\theta})\}$ in Eqn.~(\ref{eqn:obj_QAS}). Similar to conventional VQAs, QAS optimizes trainable parameters in an iterative manner. At the $t$-th iteration, QAS uniformly samples an ansatz $\bm{a}^{(t)}$ from $\mathcal{S}$ (i.e., an index list indicated by $\mathcal{A}$). To minimize $\mathcal{L}$ in Eqn.~(\ref{eqn:obj_QAS}), the parameters attached to the ansatz $\bm{a}^{(t)}$ is updated to $\bm{\theta}^{(t+1)}=\bm{\theta}^{(t)} - \eta \partial \mathcal{L}(\bm{\theta}^{(t)}, \bm{a}^{(t)}, \mathcal{Z}, \mathcal{E}_{\bm{a}^{(t)}})/\partial \bm{\theta}^{(t)}$, with $\eta$ being the learning rate. The total number of updating is set as $T$. Note that since the optimization of VQAs is NP-hard  \cite{bittel2021training},  empirical studies generally restrict $T$ to be less than $ O(poly(QNL))$ to obtain an estimation within a reasonable runtime cost.

To avoid the computational issue encountered by the two-stage optimization method, QAS leverages weight sharing strategy developed in deep neural architecture search \cite{elsken2019neural} to parameterize ansatze in $\mathcal{S}$ via a specified correlation rule.     Concretely, for any ansatz $\bm{a}'\in \mathcal{S}$, if the layout of the single-qubit gates of the $l$-th layer between $\bm{a}'$ and $\bm{a}^{(t)}$ is identical with $\forall l\in[L]$, then   $\mathcal{A}$ uses the training parameters $\bm{\theta}^{(t)}$ assigned to $U_l(\bm{\theta}^{(t)}, \bm{a}^{(t)})$ to parametrize  $U_l(\bm{\theta}^{'}, \bm{a}')$, regardless of variations in the layout of other layers. We remark that the parameterization shown above is efficient,  which can be accomplished by comparing the generated index list and the stored index lists. In addition, the above correlated updating rule implies that the parameters of unsampled ansatze are never stored the in classical memory.  To this end, even though the size of the ansatze pool exponentially scales in terms of $N$ and $L$, QAS harnesses supernet and weight sharing strategy to guarantee its applicability towards large-scale problems.

3. (Ranking.) After $T$ iterations, QAS uniformly samples $K$ ansatze from $\mathcal{S}$ (i.e., $K$ index lists generated by $\mathcal{A}$), ranks their performance, and then assigns the ansatz with the best performance as the output to estimate $\bm{a}^*$. Mathematically, denoted $\mathcal{K}$ as the set collecting the sampled $K$ ansatze, the output ansatz is
\begin{equation}\label{eqn:search-ansatz-rank} 
	\arg\min_{\bm{a}\in\mathcal{K} } \mathcal{L}(\bm{\theta}^{(T)}, \bm{a}, \mathcal{Z}, \mathcal{E}_{\bm{a}}).
\end{equation}
In QAS, $K$ is a hyper-parameter to balance the tradeoff the efficiency and performance. To avoid the exponential runtime complexity of QAS, the setting of $K$ should polynomially scale with $N$, $L$, and $Q$. Besides random sampling, other methods such as evolutionary algorithms can also be used to establish $\mathcal{K}$ with better performance. See Supplementary \ref{append:evolution} for details.

4. (Fine tuning). QAS employs the trained parameters $\bm{\theta}^{(T)}$ to fine tune the output ansatz in Eqn.~(\ref{eqn:search-ansatz-rank}).

 \begin{figure*} 
	\centering
\includegraphics[width=0.99\textwidth]{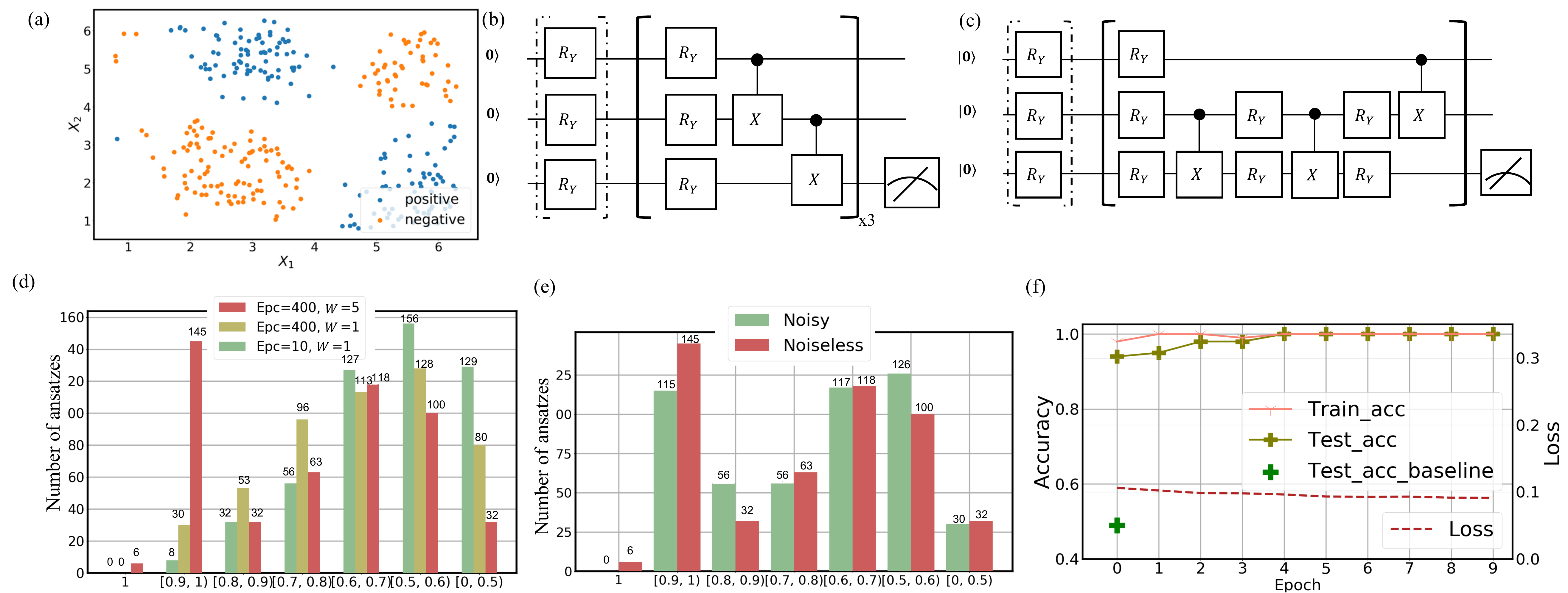}
\caption{\small{\textbf{Simulation results for the classification task.} (a) The illustration of some examples in $\mathcal{D}$ with first two features. (b) The implementation of the quantum kernel classifier for benchmarking. The quantum gates highlighted by dashed box refer to the encoding layer that transforms the classical input $\bm{x}^{(i)}$ into the quantum state. The quantum gates located in the solid box refer to $U_l(\bm{\theta})$ in Eqn.~(\ref{eqn:UL_def}) with  $L=3$.  (c) The output ansatz of QAS under the noisy setting. (d) The validation accuracy of QAS under the noiseless case. The label `Epc=a, W =b' represents that the number of epochs and supernets is  $T=a$ and $W=b$, respectively. The x-axis means that the validation accuracy of the sampled ansatz is in the range of $[c,d)$, e.g., $c=0.5$, and  $d=0.6$. (e) The comparison of QAS between the noiseless and noisy cases. The hyper-parameters setting for both cases is $T=400$, $K=500$, and $W=5$. The labeling of x-axis is identical to the subfigure (d). (f) The performance of the quantum kernel classifier (labeled by `Test\_acc\_baseline')  and QAS (labeled by `Train/Test\_acc')  at the fine tuning stage under the noisy setting. }}
\label{fig:ML_res}
\end{figure*}

\medskip 
We empirically observe fierce competition among different ansatze in $\mathcal{S}$ when optimizing QAS (See Supplementary \ref{append:ML} for details).  Namely, suppose  $\mathcal{S}$ can be decomposed into two subsets $\mathcal{S}_{\text{good}}$ and $\mathcal{S}_{\text{bad}}$, where the subset $\mathcal{S}_{\text{good}}$ ($\mathcal{S}_{\text{bad}}$) collects  ansatze in the sense that they all attain relatively good (bad) performance via independently training. For instance, in the classification task, the ansatz in $\mathcal{S}_{\text{good}}$ ($\mathcal{S}_{\text{bad}}$) promises a classification accuracy above (below) $99\%$. However, when we apply QAS to accomplish the same classification task, some ansatze in $\mathcal{S}_{\text{bad}}$ may outperform certain ansatze in $\mathcal{S}_{\text{good}}$. This observation hints the hardness of optimizing correlated trainable parameters among all ansatze accurately, where the learning  performance of a portion of ansatze in $\mathcal{S}_\text{good}$ is no better than training them independently.

To relieve fierce competition among ansatze in $\mathcal{S}$ and further boost performance of QAS, we slightly modify the initialization and optimization steps of QAS. Specifically, instead of exploiting a single supernet, QAS involves $W$ supernets to optimize the objective function in Eqn.~(\ref{eqn:obj_QAS}). The weight sharing strategy applied to  $W$ supernets are independent with each other, where the parameters corresponding to $W$ supernets are separately initialized and updated. At the training and ranking stages,  $W$ supernets separately utilize weight sharing strategy to parameterize the sampled ansatz $\bm{a}^{(t)}$  to obtain $W$ values of $\mathcal{L}(\bm{\theta}^{(t,w)}, \bm{a}^{(t)}, \mathcal{Z}, \mathcal{E}_{\bm{a}})$, where $\bm{\theta}^{(t,w)}$ refers to the parameters corresponding to the $w$-th supernet. Then, the parameters applied to the ansatz $\bm{a}^{(t)}$ is categorized into the $w'$-th supernet when $w'=\arg\min_{w\in[W]}\mathcal{L}(\bm{\theta}^{(t,w)}, \bm{a}^{(t)}, \mathcal{Z}, \mathcal{E}_{\bm{a}})$.

We last emphasize how QAS enhances the learning performance of hardware-efficient ansatz  $U(\bm{\theta})$ in Eqn.~(\ref{eqn:UL_def}).  Recall that the central aim of QAS is to seek a good ansatz associated with optimized parameters to minimize $\mathcal{L}(\bm{\theta}, \bm{a}, \mathcal{Z},\mathcal{E}_{\bm{a}})$ in Eqn.~(\ref{eqn:obj_QAS}). In other words, given   $U=\prod_{l=1}^LU_l(\bm{\theta})$, a good ansatz is located by dropping some unnecessary multi-qubit gates and substituting single-qubit gates in $U_l(\bm{\theta})$ for $\forall l \in [L]$. Following this routine, several studies have proved that removing multi-qubit gates to reduce the entanglement of the ansatz contributes to alleviate barren plateaus \cite{marrero2020entanglement,patti2020entanglement}.  In addition, a recent study \cite{Tobias2021capacity} unveiled that the choice of the quantum circuit architecture can significantly affect the expressive power of the ansatz and the learning performance. Since the objective function of QAS implicitly evaluates the effect of different ansatze, our proposal can be employed as a powerful tool to enhance the learning performance of VQAs.   Refer to Method for  further explanation about the role of supernet, weight sharing, and analysis of the memory cost and runtime complexity of QAS.

\medskip
\noindent\textbf{Simulation and experimental results.}  The proposed QAS is universal and facilitates a wide range of VQAs based learning tasks, e.g., machine learning \cite{huang2021power,du2018implementable,cong2019quantum,wang2021towards}, quantum chemistry \cite{peruzzo2014variational,google2020hartree}, and quantum information processing \cite{larose2019variational,du2019efficient}. In the following, we separately apply QAS to accomplish a classification task and a variational quantum eigen-solver (VQE) task to confirm its capability towards the performance enhancement. All numerical simulations are implemented in Python in conjunction with the PennyLane and the Qiskit packages \cite{bergholm2018pennylane,Qiskit}. Specifically, PennyLane is the backbone to implement QAS and Qiskit supports different types of noisy models. We defer the explanation of basic terminologies in  machine learning and quantum chemistry in Appendices  \ref{append:ML} and \ref{append:q_chem}.

Here we first apply QAS to achieve a binary classification task  under both the noiseless and noisy scenarios. Denote  $\mathcal{D}$ as the  synthetic dataset, where its construction rule follows the proposal of the quantum kernel classifier \cite{havlivcek2019supervised}.  The  dataset $\mathcal{D}$ contains $n=300$ samples. For each example $\{\bm{x}^{(i)},y^{(i)}\}$, the feature dimension of the input  $\bm{x}^{(i)}$ is $3$ and the corresponding label $y^{(i)}\in\{0, 1\}$ is binary. Examples of $\mathcal{D}$ are shown in Figure \ref{fig:ML_res}.  At the data preprocessing stage, we split the dataset $\mathcal{D}$ into the training set $\mathcal{D}_{tr}$, validation set $\mathcal{D}_{va}$, and test set $\mathcal{D}_{te}$ with size $n_{tr}=100$, $n_{va}=100$, and $n_{te}=100$.  The explicit form of the objective function is 
 \begin{equation}\label{eqn:loss_ML}
 	\mathcal{L} = \frac{1}{n_{tr}}\sum_{i=1}^{n_{tr}}\left(\tilde{y}^{(i)}(\mathcal{A},\bm{x}^{(i)},\bm{\theta})-y^{(i)}\right)^2,
 \end{equation} 
 where $\{\bm{x}^{(i)}, y^{(i)}\}\in \mathcal{D}_{tr}$ and $\tilde{y}^{(i)}(\mathcal{A},\bm{x}^{(i)},\bm{\theta})\in[0, 1]$ is the output of the quantum classifier (i.e., a function taking  the input $\bm{x}^{(i)}$, the supernet $\mathcal{A}$, and the trainable parameters $\bm{\theta}$). The  training (validation and test) accuracy is measured by $\sum_{i}\mathbbm{1}_{g(\tilde{y}^{(i)})=y^{(i)}}/n_{tr}$ ($\sum_{i}\mathbbm{1}_{g(\tilde{y}^{(i)})=y^{(i)}}/n_{va}$ and $\sum_{i}\mathbbm{1}_{g(\tilde{y}^{(i)})=y^{(i)}}/n_{te}$) with $g(\tilde{y}^{(i)})$ being the predicted label for $\bm{x}^{(i)}$. We also apply the quantum kernel classifier  proposed by \cite{havlivcek2019supervised} to learn $\mathcal{D}$ and compare its  performance with QAS, where the implementation of such a quantum classifier is shown in Figure \ref{fig:ML_res} (b). See Supplementary \ref{append:ML} for more discussion about the construction of $\mathcal{D}$ and the employed quantum kernel classifier. 
 
 The hyper-parameters  for QAS are as follows.  The number of supernets is $W=1$ and $W=5$, respectively. The circuit depth for all supernets is set as $L=3$.  The search space of QAS is formed by  two types of quantum gates. Specifically, at each layer $U_l(\bm{\theta})$, the parameterized gates are fixed to be the rotational quantum gate along $Y$-axis $\RY$. For the two-qubit gates, denoted the index of three qubits as $(0, 1, 2)$, QAS explores whether applying CNOT gates to the qubits pair $(0,1)$, $(0,2)$, $(1,2)$ or not. Hence, the size of $\mathcal{S}$ equals to $|\mathcal{S}|=8^3$. The number of sampled ansatze for ranking is set as $K=500$,  The setting $K\approx |\mathcal{S}|$, enables us to understand how the number of supernets $W$, the number of epochs $T$, and the system noise effect the learning performance of different ansatze in the ranking stage.

\begin{figure*}[ht!]
	\centering
\includegraphics[width=0.99\textwidth]{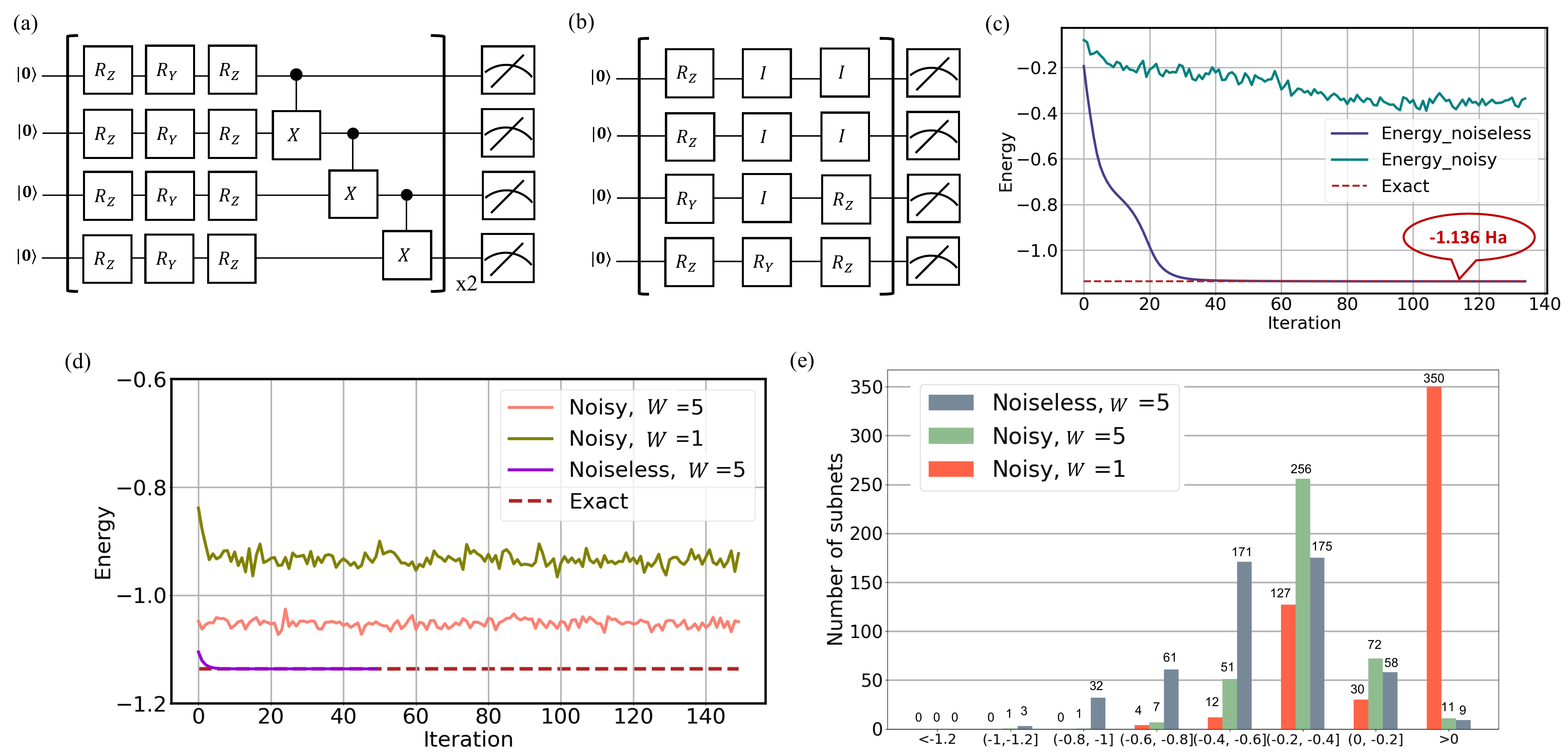} 
\caption{\small{\textbf{Simulation results for the ground state energy estimation of Hydrogen.} (a) The implementation of the conventional VQE. (b) The output ansatz of QAS under the noisy setting. (c) The training performance of VQE under the noisy and noiseless settings. The label `Exact' refers to the accurate result $E_m$. (d) The  performance of the output ansatz of QAS under both the noisy and noiseless settings. (e) The performance of QAS at the ranking state. The label `W=b' refers to the number of supernets, i.e., $W=b$. The x-axis means that the estimated energy of the sampled ansatz is in the range of $(c,d]$, e.g., $c=-0.6 \Ha$, and  $d=-0.8\Ha$.  }}
\label{fig:chem_res} 
\end{figure*}

 Under the noiseless scenario, the performance of QAS with three different settings is exhibited in Figure \ref{fig:ML_res} (d). In particular, QAS with $W=1$ and $T=10$ attains the worst performance, where the validation accuracy for most ansatze concentrates on $50\%-60\%$, highlighted by the green bar. With increasing the number of epochs to $T=400$ and fixing $W=1$, the performance is slightly improved, i.e., the number of ansatze that achieves validation accuracy above $90\%$ is $30$, highlighted by the yellow bar. When $W=5$ and $T=400$, the performance of QAS is dramatically enhanced, where the validation accuracy of $151$ ansatze is above $90\%$. The comparison between the first two settings indicates the correctness of utilizing QAS to accomplish VQA-based learning tasks in which QAS learns useful feature information and achieves better performance with respect to the increased epoch number $T$.  The varied performance of the last two settings reflects the fierce competition phenomenon among ansatze and validates the feasibility to adopt $W>1$ to boost performance of QAS. We retrain the output ansatz of QAS under the setting: $W=5$ and $T=400$, both the training and test accuracies converge to $100\%$ within $15$ epochs, which is identical to the original quantum kernel classifier.

The performance of the original quantum kernel classifier is evidently degraded when the depolarizing error for the single-qubit and two-qubit gates is set as $0.05$ and $0.2$, respectively. As shown in the lower plot of Figure \ref{fig:ML_res} (f), the training and test accuracies of the original quantum kernel classifier drop to $50\%$ (almost conduct a random guess) under the noisy setting. The degraded performance is caused by the  large amount of accumulated noise, where the classical optimizer fails to receive the valid optimization information. By contrast, QAS can achieve a good performance under the same noise setting. As shown in Figure \ref{fig:ML_res} (e), with setting $W=5$ and $T=400$, the validation accuracy of  $115$ ansatze is above $90\%$ under the noisy setting. The ansatz that attains the highest validation accuracy is shown in \ref{fig:ML_res} (c). Notably, compared with the original quantum kernel classifier in Figure \ref{fig:ML_res} (b), the searched ansatz contains fewer CNOT gates. This implies that, under the noisy setting formulated above, QAS suppresses  the noise effect and improves the training performance by adopting few CNOT gates.  When we retrain the obtained ansatz with $10$ epochs, both the train and test accuracies achieve $100\%$, as shown in the upper plot of Figure \ref{fig:ML_res} (f). These results indicate the feasibility to apply QAS to achieve the noise inhibition  and trainability enhancement. 

We defer the omitted simulation results and the exploration of fierce competition to Supplementary \ref{append:ML}. In particular, we assess the learning performance of the quantum classifier with the hardware-efficient ansatz and the ansatz searched by QAS under the noise model extracted from the real quantum device, i.e., `Ibmq\_lima'. The achieved simulation result indicates that the ansatz obtained by QAS outperforms the conventional quantum classifier.
 
 We next apply QAS to find the ground state energy of the Hydrogen molecule \cite{o2016scalable,kandala2017hardware} under both the noiseless and noisy scenarios. The molecular hydrogen Hamiltonian is formulated as
 \begin{eqnarray}\label{eqn:ground_H2}
 	&& H_h =  g + \sum_{i=0}^3 g_i Z_i + \sum_{i=1,k=1,i<k}^3 g_{i,k} Z_i Z_k  + g_{a} Y_0X_1X_2Y_3  \nonumber\\
 	&& + g_{b}Y_0Y_1X_2X_3 + g_{c}X_0X_1Y_2Y_3  + g_{d}X_0Y_1Y_2X_3,
 \end{eqnarray}
where $\{X_i, Y_i, Z_i\}$ denote the Pauli matrices acting on the $i$-th qubit and the real scalars $g$ with or without subscripts are efficiently computable functions of the hydrogen-hydrogen bond length (see Supplementary \ref{append:q_chem} for details about $H_h$ and $g$). The ground state energy calculation amounts to computing the lowest energy eigenvalues of $H_h$, where the accurate value is $E_m = -1.136 \Ha$    \cite{bergholm2018pennylane}.  To tackle this task, the conventional  VQE  \cite{peruzzo2014variational}  and its variants \cite{wang2019accelerated,stokes2020quantum,mitarai2019generalization}  optimize the trainable parameters in  $U(\bm{\theta})$ to prepare the ground state $\ket{\psi^*}=U(\bm{\theta}^*)\ket{0}^{\otimes 4}$ of $H_h$, i.e., $E_m=\braket{\psi^*|H_h|\psi^*}$. The implementation of $U(\bm{\theta})$ is illustrated in Figure \ref{fig:chem_res} (a). Under the noiseless setting, the estimated energy of VQE fast converges to the target result $E_m$ within 40 iterations, as shown in Figure \ref{fig:chem_res} (c).  

The hyper-parameters of QAS to compute the lowest energy eigenvalues of $H_h$ are as follows. The number of supernets has two settings, i.e.,  $W=1$ and $W=5$, respectively. The layer number for all ansatze is $L=3$. The number of iterations and  sampled ansatze for ranking is $T=500$ and $K=500$, respectively.  The search space of QAS for the single-qubit gates is fixed to be the rotational quantum gates along $Y$ and $Z$-axis.  For the two-qubit gates, denoted the index of four qubits as $(0, 1, 2, 3)$, QAS explores whether applying CNOT gates to the qubits pair $(0,1)$, $(1,2)$, $(2,3)$ or not. Therefore, the total number of ansatze equals to $|\mathcal{S}|=128^3$. The performance of QAS with $W=5$ is shown in Figure \ref{fig:chem_res} (d). Through retraining the obtained ansatz of QAS with $50$ iterations, the estimated energy converges to $E_m$, which is the same with the conventional VQE.   

The performance between the conventional VQE and QAS is largely distinct when the noisy model described {in the classification task is deployed.  Due to the large amount of gate noise, the estimated ground energy of the conventional VQE converges to $-0.4\Ha$, as shown in Figure \ref{fig:chem_res} (c). In contrast, the estimated ground energy of QAE with $W=1$ and $W=5$ achieves $-0.93\Ha$ and $-1.05\Ha$, respectively. Both of them are closer to the target result $E_m$ compared with the conventional VQE. Moreover, as shown in Figure \ref{fig:chem_res} (e),  a lager $W$ implies better performance of QAS, since the estimated energy of most ansatze is below $-0.6 \Ha$ when $W=5$, while the estimated energy of $350$ ansatze is above $0 \Ha$ when $W=1$. We illustrate the generated ansatz of QAS with $W=5$ in Figure \ref{fig:chem_res} (b). In particular, to mitigate the effect of gate noise, this generated ansatz does not contain any CNOT gate, which is applied to a very large  noise level. Recall that a central challenge in quantum computational chemistry is whether NISQ devices can outperform classical methods already available \cite{mcardle2020quantum}.  The achieved results in QAS can provide a good guidance to answer this issue. Concretely, the searched ansatz in Figure \ref{fig:chem_res}, which only produces the separable states that can be efficiently simulated by classical devices, suggests that VQE method may not outperform classical methods when NISQ devices contain large gate noise.   

Note that more simulation results are deferred to Supplementary. Specifically, in Supplementary \ref{append:q_chem}, we exhibit more results of the above task. Furthermore, we implement VQE with the hardware-efficient ansatz and the ansatz searched by QAS on the real superconducting quantum hardware, i.e., `Ibmq\_ourense', to estimate the ground state energy of $H_h$. Due to the runtime issue, we complete the optimization and ranking using the classical backend and perform the final runs on the IBMQ cloud. Experimental result indicates that the ansatz obtained by QAS outperforms the conventional VQE, where the estimated energy of the former is $-0.96\Ha$ while the latter is $-0.61\Ha$. Then, in Supplementary \ref{append:evolution}, we exhibit that utilizing the evolutionary algorithms to establish $\mathcal{K}$ can dramatically improve the performance of QAS. Subsequently, in  Supplementary \ref{appendix:barren plateaus},  we provide the numerical evidence that QAS  can alleviate the influence of barren plateaus. Last, we present a variant of QAS to  tackle large-scale problems with the enhanced performance in Supplementary \ref{append:pro-QAS}.

\medskip  
\noindent\textbf{DISCUSSION}\\
In this study, we devise QAS to dynamically and automatically design ansatz for VQAs. Both simulation and experimental results validate the effectiveness of QAS.  Besides  good performance, QAS only requests  similar computational resources with conventional VQAs with fixed ansatze and is compatible with all quantum systems. Through incorporating QAS with other advanced error mitigation and trainability enhancement techniques, it is possible to seek more applications that can be realized on NISQ machines with potential advantages. 

 There are many critical questions remaining in the study of QAS. Our future work includes the following several directions. First, we will explore better strategies to sample ansatz at each iteration. For example, the reinforcement learning techniques, which is used to construct optimal sequences of unitaries to accomplish quantum simulation tasks \cite{yao2020reinforcement}, may contribute to this goal.  Next, we will design a more advanced strategy to shrink the parameter space  while not degrading the learning performance. Subsequently, to further boost the performance of QAS, we will leverage some prior information of the learning problem such as the symmetric property and some post-processing strategies that remove redundant gates of the searched ansatz.  In addition, we will delve to theoretically understanding the fierce competition.  In the end, it is intriguing to explore applications of QAS beyond VQAs such as optimal quantum control and the approximation of the target unitary using the limited quantum gates.

\medskip
\noindent{\textbf{Methods}}\\
 
\noindent{\textbf{M.1 The classical analog of QAS}}

The classical analog of the learning problem in  Eqn.~(\ref{eqn:obj_QAS}) is the neural network architecture search \cite{elsken2019neural}. Recall that the success of deep learning is largely attributed to  novel neural architectures for specific learning tasks, e.g., the convolutional neural networks for image processing tasks \cite{goodfellow2016deep}. However, deep neural networks designed by human experts are generally time-consuming and error-prone \cite{elsken2019neural}. To tackle this issue, the neural architecture search approach, i.e., the process of automating architecture engineering, has been widely explored, and achieved state of the art performances in many learning tasks   \cite{pham2018efficient,huang2021greedynasv2,liu2018progressive,you2020greedynas,yang2020ista}. Despite having a similar aim, naively generalizing classical results to the quantum scenario to accomplish Eqn.~(\ref{eqn:obj_QAS}) is infeasible due to the distinct basic components: neurons versus quantum gates, classical correlation versus entanglement, the barren plateau phenomenon, the quantum noise affect, and physical hardware restrictions. These differences and extra limitations further intensify the difficulty of searching the optimal quantum circuit architecture $\bm{a}^*$, compared with the classical setting. In the following, we explain the omitted implementation details of QAS.
 
\medskip
\noindent{\textbf{M.2 Weight sharing strategy}}
 
The role of weight sharing strategy is reducing the parameter space to enhance the learning performance of QAS within a reasonable runtime and memory usage. Intuitively, this strategy  correlates parameters among different ansatze in $\mathcal{S}$ based on a specified rule. In this way, we can jointly optimize $(\bm{\theta},\bm{a})$ to estimate $(\bm{\theta}^*,\bm{a}^*)$, where the updated parameters for one ansatz can also enhance the learning performance of other ansatze when the correlation criteria is satisfied. As explained in  Figure \ref{fig:weight-share}, weight sharing strategy adopted in QAS squeezes the parameter space from $O(dQ^{NL})$ to $O(dLQ^{N})$. Meantime, our simulation results indicate that the reduction of parameter space enables QAS to achieve a good performance  within a reasonable runtime complexity.

We remark that through  adjusting the correlation criteria applied to weight sharing strategy, the parameter space can be further reduced. For instance, when all parameters in an ansatz are correlated, the size of the parameter space reduces to $O(1)$. With this regard, another feasible correlation rule for QAS is unifying the single-qubit gates for all ansatze as $U_3=\RZ(\alpha)\RY(\beta)\RZ(\gamma)$. In other words, QAS only adjusts the arrangement of two-qubit gates to enhance the learning perforamnce. From the practical perspective, this setting is reasonable since the gate error introduced by the single-qubit gates is much less than that of two-qubit gates. }

 \begin{figure}
	\centering
\includegraphics[width=0.48\textwidth]{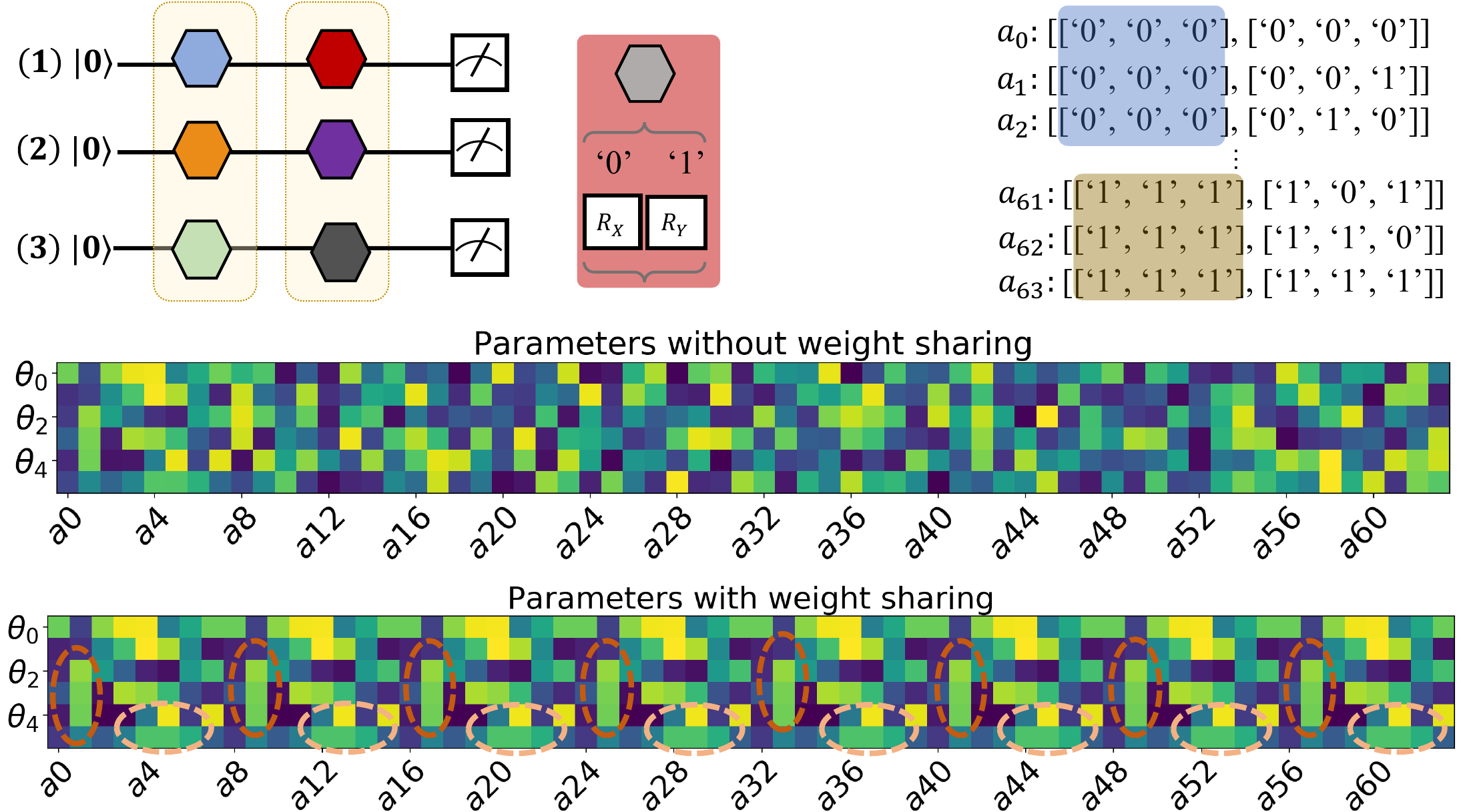}
\caption{\small{\textbf{A visualization of weight sharing strategy.} The upper left panel depicts the potential ansatze when $N=3$, $L=2$, and the choices of quantum gates are $\{\RX, \RY\}$ with $Q=2$. The total number of ansatze is $Q^{NL}=64$. The upper right panel illustrates how to use the indexing technique to accomplish the weight sharing. The label `$a_{i}$' refers to the ansatz $\bm{a}^{(i)}$. Namely, for any two ansatze, if the indexes in the $l$-th array are identical (highlighted by the blue and brown regions), then their trainable parameters in the $l$-th layer are the same. The two heat-maps demonstrated in the lower panel visualize the trainable parameters of $64$ ansatze. The label `$\bm{\theta}_i$' refers to the parameter assigned to the $i$-th rotational quantum gate. Note that when weight sharing strategy is applied, the trainable parameters are reused for different ansatze, as indicated by the dashed circles.  }}
\label{fig:weight-share}
\end{figure}
 
 \medskip 
\noindent \textbf{M.3 Supernet}  

We next elucidate supernet used in QAS. As explained in the main text, supernet has two important roles, which are  constructing the ansatze pool $\mathcal{S}$ and  parameterizing each ansatz in $\mathcal{S}$ via the specified weight sharing strategy. In other words, supernet defines the search space, which subsumes all candidate ansatze, and the candidate ansatze in $\mathcal{S}$ are evaluated through inheriting weights from the supernet.  Rather than training numerous separate ansatze from scratch, QAS trains supernet just once (Step 2 in Figure \ref{fig:QAS}), which significantly cuts down the search cost.

We next explain how QAS leverages the indexing technique to construct $\mathcal{S}$ when the available quantum gates include both single-qubit and two-qubit gates. Following  notation in the main text, suppose that $N=5$, $L=1$, and the choices of single-qubit gates and two-qubit gates are $\{\RY,\RZ\}$  and $\{\CNOT, \mathbb{I}_4\}$, respectively. In QAS, supernet $\mathcal{A}$ indexes $\{\RY$, $\RZ$, $\CNOT$, $\mathbb{I}_4\}$ as $\{`0$', `$1$', `T', `F'\}. Moreover, we suppose that the topology of the deployed quantum machine yields a chain structure, i.e., $Q1 \leftrightarrow Q2 \leftrightarrow Q3\leftrightarrow Q4\leftrightarrow Q5 $. With setting  $a,b,c,d,e\in$ \{`0', `1'\} and $A,B,C,D\in$ \{`T', `F'\}, the index list [`a', `b', `c', `d', `e', `A', `B', `C', `D'] tracks  all candidate ansatze in  $\mathcal{S}$, e.g., [`0', `0', `0', `0', `0', `T', `T', `T', `T'] describes the ansatz $(\prod_{i=1}^4\CNOT_{i,i+1})(\otimes_{i=1}^5 \RY(\bm{\theta}_i))$  and [`1', `1', `1', `1', `1', `F', `F', `F', `F'] describes the ansatz $\otimes_{i=1}^5 \RZ(\bm{\theta}_i)$.
 
 \medskip
\noindent \textbf{M.4 Memory cost and runtime complexity}
  
We first analyze the runtime complexity of QAS.  In particular, at the first step, the setup of supernet, i.e., configuring out the ansatze pool and the correlating rule, takes $O(1)$ runtime. In the second step,  QAS proceeds $T$ iterations to optimize trainable parameters. The runtime cost of QAS at each iteration scales with $O(d)$, where $d$ refers to the number of trainable parameters in Eqn.~(\ref{eqn:exp_erm}). Such cost origins from the calculation of gradients via parameter shift rule, which is similar with the optimization of VQAs with a fixed ansatz. To this end, the total runtime cost of the second step is $O(dT)$. In the ranking step, QAS   samples $K$ ansatze and compares their objective values using the optimized parameters. This step takes at most $O(K)$ runtime. In the last step, QAS fine tunes the parameters based on the searched ansatz with few iterations (i.e., a very small constant). The required runtime is identical to conventional VQAs, which satisfies $O(d)$. The total runtime complexity of QAS is hence $O(dT+K)$.

We next analyze the memory cost of QAS. Specifically, the first step requests $O(QNL)$ memory to specify the ansatze pool via the indexing technique. Recall the memory cost in this step is dominated by configuring the index space, which requests at most $O(QNL)$ memory. This is because in the worst case, the allowed $Q$ choices of quantum gates for the varied qubit at the varied layer are exactly different. To store  information that describes choices of gates for different qubits at different position, the memory cost scales with $O(QNL)$. In the second step, QAS totally outputs $T$ index lists corresponding to the architecture of $T$ ansatze. This requires at most $O(TNL)$ memory cost. Moreover, QAS explicitly updates at most $Td$ parameters (we omit those parameters that are implicitly updated via weight sharing strategy, since they do not consume the memory cost). To this end, the memory cost of the second step is $O(TNL+Td)$. In the third step, QAS samples $K$ index lists that describe  the circuit architecture of $K$ ansatze. This requires at most $O(KNL)$ cost. Moreover, according to weight sharing strategy, the memory cost of storing the corresponding parameters is $O(Kd)$.  The memory cost of the last step is identical to the conventional VQAs with a fixed ansatz, which is $O(d)$.  The total memory cost of QAS is hence $O(Td+TNL+Kd)$.

To better understand how the  computational complexity scales with $N$, $L$ and $Q$, in the following, we set the total number of iterations in Step 2 and the number of sampled ansatze in Step 3 as $T=O(QNL)$  and $K=O(QNL)$, respectively. Note that since the size of $\mathcal{S}$ becomes indefinite, it is reasonable to set $K$ as $O(QNL)$ instead of a constant used in the numerical simulations. Under the above settings, we conclude that the runtime complexity and  the memory cost of QAS are  $O(dQNL)$ and $O(dQNL+QN^2L^2)$, respectively.
 
 We remark that when $W$ supernets are involved, the required memory cost and runtime complexity of QAS linearly scales with respect to $W$. Moreover,  employing adversarial bandit learning techniques \cite{bubeck2012regret} can exactly remove this overhead (See Supplementary \ref{appedn:bandit} for details).

\medskip
\noindent\textbf{DATA AVAILABILITY}\\
The datasets generated and/or analyzed during the current study are  available from Y.D. on reasonable request. 

\medskip
\noindent\textbf{CODE AVAILABILITY}\\
The source code of QAS to reproduce all numerical experiments is available on the GitHub repository \url{https://github.com/yuxuan-du/Quantum_architecture_search/}.

\medskip
\noindent
\textbf{AUTHOR CONTRIBUTIONS}\\
Y.D. and D.T. conceived this work. Y.D, S.Y., and M.H. accomplished the theoretical analysis. Y.D. and T.H. conducted numerical simulations. All authors reviewed and discussed the analysis and results, and contributed towards writing the manuscript. Correspondence to M.H. and D.T..

\medskip
\noindent
\textbf{COMPETING INTERESTS}\\
The authors declare no competing interests.

%

\newpage

\appendix
\onecolumngrid
\newpage

We organize the Supplementary as follows.  In Supplementary \ref{appedn:bandit}, we establish the connection between the bandit learning and the ansatz assignment task  and discuss how to exploit bandit learning algorithms to further advance the ansatz assignment task.  We then provide explanations and  simulation results related to the fierce  competition phenomenon and the classification task in Supplementary \ref{append:ML}.  Afterwards, we  present simulation and experiment details about the quantum chemistry tasks in Supplementary \ref{append:q_chem}. Subsequently, we exhibit how to introduce evolutionary algorithms into the ranking stage to boost the performance of QAS in Supplementary \ref{append:evolution}. Next, we empirically explore the trainability of QAS through the lens of barren plateaus in Supplementary \ref{appendix:barren plateaus}. Last, we demonstrate a variant of QAS to effectively accomplish large-scale problems in Supplementary \ref{append:pro-QAS}.

\section{The ansatz assignment task}\label{appedn:bandit}
 In this section, we first connect the ansatz assignment task with the  adversarial bandit learning problem. We then compare the method used in QAS with all bandit algorithms in terms of the regret measure. We last explain how to employ advanced bandit learning algorithms to reduce the runtime complexity of the ansatz assignment task.

\subsection{The connection between the adversarial bandit learning and the ansatz assignment}\label{append:sub_bandit_1}
Let us first introduce the adversarial bandit learning. In the adversarial bandit learning \cite{bubeck2012regret}, a player has $W$ possible arms to choose. Denote the total number of iterations as $T$. At the $t$-th iteration,
\begin{itemize}\label{prob:adver_bandit}
	\item  The player chooses an arm  $w^{(t)}\in[W]$ with a deterministic strategy or sampling from a certain distribution $\mathcal{P}_w$;
	\item The adversary chooses a cost $c^{(t)}(w^{(t)})$ for the chosen arm $w^{(t)}$;
	\item The cost of the selected arm $w^{(t)}$, i.e., $c^{(t)}(w^{(t)})$ with $w^{(t)} \in [W]$, is revealed to the player.
\end{itemize}   
The goal of the adversarial bandit learning is minimizing the total cost over $T$ iterations, where its performance is quantified by the regret $r_T$, i.e., 
 \begin{equation}\label{eqn:appd_thm_1_1}
 	r_T = \sum_{t=1}^{T} c^{(t)}(w^{(t)}) - \min_{w  \in[W] }\sum_{t=1}^T  c^{(t)}(w).
 \end{equation}
 Intuitively, the regret $r_T$ compares the  cumulative cost of the selected arms $\{w^{(t)}\}_{t=1}^T$ with the best arm in hindsight.   If $r_T = o(T)$, where the regret can be either negative or scales at most sublinearly with $T$, we say that the player is learning;  otherwise, when $r_T= \Theta(T)$ such that the regret scales linearly with $T$,  we say that the player is not learning, since the averaged cost per-iteration does not decrease with time.

We now utilize the language of the adversarial bandit learning to restate the ansatz assignment problem. In QAS, each arm refers to a supernet and the number of arms equals to the number of supernets. The cost $c^{(t)}(w^{(t)})$ is equivalent to the objection function $\mathcal{L}(\bm{\theta}^{(t, w)}, \bm{a}^{(t)})$ in Eqn.~(\ref{eqn:obj_QAS}), where  $\bm{a}^{(t)}$ refers to the sampled ansatz $\bm{a}^{(t)}\in \mathcal{S}$, and $\bm{\theta}^{(t, w)}$ represents the trainable parameters of  the $w$-th supernet $\mathcal{A}^{(w)}$. The aim of the ansatz assignment is  to allocate $\{\bm{a}^{(t)}\}_{t=1}^T$ to the best sequence of arms  (supernets)  to minimize the cumulative cost. Denote the selected sequence of arms (indices of supernets) of QAS as $\{I_w^{(t)}\}_{t=1}^T$. The regret  in Eqn.~(\ref{eqn:appd_thm_1_1}) can be rewritten as 
\begin{equation}\label{eqn:append1_QAS_regrt}
	R_T =  \sum_{t=1}^T \mathcal{L}(\bm{\theta}^{(t,I_w^{(t)})}, \bm{a}^{(t)})  -\min_{w  \in[W] }\sum_{t=1}^T\mathcal{L}(\bm{\theta}^{(t,w)}, \bm{a}^{(t)}).
\end{equation}  

\subsection{The comparison between the strategy used in QAS and all bandit algorithms}
The following theorem shows that the strategy used in QAS outperforms all bandit algorithms in terms of the regret measure.    
\begin{thm}\label{thm1}
Let $W$ and $T$ be the number of supernets and iterations, respectively. Suppose that the ansatz $\bm{a}^{(t)}$ is assigned to the $I_w^{(t)}$-th supernet $\mathcal{A}^{(I_w^{(t)})}$ with $I_w^{(t)}\in[W]$ at the $t$-th iteration, where the corresponding  objective function in Eqn.~(\ref{eqn:obj_QAS}) is $\mathcal{L}(\bm{\theta}^{(t,I_w^{(t)})},\bm{a}^{(t)}) \in[0, 1]$. Define the regret as 
\begin{equation}\label{eqn:regret_thm1} 
	R_T =  \sum_{t=1}^T \mathcal{L}(\bm{\theta}^{(t,I_w^{(t)})}, \bm{a}^{(t)})  -\min_{w \in [W]}\sum_{t=1}^T\mathcal{L}(\bm{\theta}^{(t,w)}, \bm{a}^{(t)}),
\end{equation}  
where the randomness is over the selection of $I_w^{(t)}$. The method used in QAS to determine $\{I_w^{(t)}\}$ promises the regret $R_T\leq 0$, while the regret for the best bandit algorithms is lower bounded by  $R_T= \Omega(T)$.    
\end{thm}

The proof of Theorem \ref{thm1} exploits the following lemma. 
\begin{lem}[Theorem 1,  \cite{gerchinovitz2016refined}]\label{lem:low_b_advb}
	Suppose $W\geq 2$ and $\delta\in (0, 1/4)$ and $T\geq 32(W-1)\log(2/\delta)$, then there exists a sequence of data $\{\bm{\theta}^{(t,I_w^{(t)})}, \bm{a}^{(t)}\}_{t=1}^T$, or equivalently, the objective values $\{\mathcal{L}(\bm{\theta}^{(t,I_w^{(t)})}, \bm{a}^{(t)})\}_{t=1}^T$ such that the regret in Eqn.~(\ref{eqn:append1_QAS_regrt}) follows  
	\begin{equation}
		\mathcal{P}\left(R_T \geq \frac{1}{27}\sqrt{(W-1)T\log(1/(4\delta))} \right) \geq \delta /2.
	\end{equation}   
\end{lem}
The lower bound given in Lemma \ref{lem:low_b_advb} indicates that under the adversarial setting,  there does not exist an adversarial bandit algorithm can achieve the regret smaller than $\Omega(\sqrt{WT\log(1/\delta)})$ with probability at least $1-\delta$.

We are now ready to prove Theorem \ref{thm1}.
\begin{proof}[Proof of Theorem \ref{thm1}]
Here we first prove the regret $R_T$ in Eqn.~(\ref{eqn:regret_thm1}) for the assignment strategy employed in QAS. We then quantity the lower bound of $R_T$ for all adversarial bandit algorithms.

Recall the assignment strategy used in QAS. Given the sampled ansatz $\bm{a}^{(t)}\in\mathcal{S}$, QAS feeds this ansatz into $W$ supernets and compares $W$ values of objective functions, i.e., $\{\mathcal{L}(\bm{\theta}^{(t,w)}, \bm{a}^{(t)})\}_{w=1}^W$. Then, the ansatz $\bm{a}^{(t)}$ is assigned to the $I_w^{(t)}$-th supernet as
	\begin{equation}\label{eqn:thm1_app_1}
		I_w^{(t)} = \arg\min_{w=1,...,W} \mathcal{L}(\bm{\theta}^{(t,w)}, \bm{a}^{(t)}). 
	\end{equation}
	
Denote the regret $R_T$ in Eqn.~(\ref{eqn:regret_thm1}) obtained by QAS as $R_T^Q$.  By exploiting the explicit definition of $I_w^{(t)}$ in Eqn.~(\ref{eqn:thm1_app_1}),  the regret $R_T^Q$  yields 
\begin{eqnarray}\label{eqn:thm1_app_2}
	R_T^Q && = \sum_{t=1}^T \mathcal{L}(\bm{\theta}^{(t,I_w^{(t)})}, \bm{a}^{(t)})  -\min_{w \in [W]}\sum_{t=1}^T\mathcal{L}(\bm{\theta}^{(t,w)}, \bm{a}^{(t)}) \nonumber\\
	&& =\sum_{t=1}^T \min_{w=1,...,W} \mathcal{L}(\bm{\theta}^{(t,w)}, \bm{a}^{(t)}) -\min_{w \in [W] }\sum_{t=1}^T\mathcal{L}(\bm{\theta}^{(t,w)}, \bm{a}^{(t)})\nonumber\\
	&& \leq 0, 
\end{eqnarray}
where the last inequality employs the fact that  the summation of minimum values of functions is less than the minimum value of summation of functions (i.e., $ \sum_t \min_x  f_t(x) \leq \min_x \sum_t f_t(x)$ and the equality is hold when the minimum of all functions $\{f_t(x)\}$ is identical). 

Denote the regret $R_T$ in Eqn.~(\ref{eqn:regret_thm1}) obtained by a given bandit algorithm as $R_T^B$. Due to Lemma \ref{lem:low_b_advb}, we achieve 
	\begin{equation}\label{eqn:thm1_append1_3}
		\mathcal{P}\left(R_T^B  \geq  \frac{1}{27}\sqrt{(W-1)T\log(1/(4\delta))} \right) \geq \delta /2.
	\end{equation}  
 In other words, for the ansatz assignment task, the regret for all adversarial bandit algorithms is lower bounded by    $R_T^B \geq \Omega(\sqrt{WT\log(1/\delta)})$ with probability $\delta$.
 
Based on Eqn.~(\ref{eqn:thm1_app_2}) and Eqn.~(\ref{eqn:thm1_append1_3}), we conclude that with high probability, no bandit learning algorithm can achieve a lower  regret than that of the strategy adopted in QAS. 
\end{proof}

\subsection{Applying bandit learning algorithms to the ansatz assignment task}
 Here we discuss how to apply bandit learning algorithms to improve the ansatz assignment task in terms of the runtime cost.   Recall the  ansatz assignment strategy used in QAS. At each iteration, the sampled ansatz should feed into $W$ supernets separately and then compare the returned $W$ objective values. In this way, the runtime complexity becomes expensive for a large $W$, as discussed in Method. The adversarial bandit learning algorithms are a promising solution to tackle the runtime issue. As explained in Supplementary \ref{append:sub_bandit_1}, when adversarial bandit learning algorithms are employed, the ansatz is only required to feed into one supernet at each iteration, while the price is inducing a relatively large regret bound.

\section{The synthetic dataset classification task}\label{append:ML}
The outline of this section is as follows.  In Supplementary \ref{append:ML_sub0}, we first introduce some basic terminologies in machine learning to make our description self-consistent. In Supplementary \ref{append:ML_sub1}, we  explain how to construct the synthetic dataset $\mathcal{D}$.  In Supplementary  \ref{append:ML_sub2}, we provide the simulation results omitted in the main text and elaborate on the fierce competition phenomenon. Last, in Supplementary \ref{appendix:subsec:QAS-ibmq-classifier},  we compare the learning performance of the quantum classifier with the hardware-efficient ansatz and the ansatz searched by QAS under the noise model extracted from the real quantum device, i.e., an IBM's 5-qubit quantum machine nameds as `Ibmq\_lima'. 

\subsection{Basic  terminologies in  machine learning}\label{append:ML_sub0}
When we apply QAS to accomplish the classification task, the terminology `epoch', which is broadly used in the field of machine learning \cite{goodfellow2016deep}, is employed to replace `iteration'. Intuitively, an epoch means that an entire dataset is passed forward through the quantum learning model. For the quantum kernel classifier used in the main text, each training example in $\mathcal{D}_{tr}$ is  fed into the quantum circuit in sequence to acquire the predicted label. Since $\mathcal{D}_{tr}$ includes in total $100$ examples, it will take $100$ iterations to complete one epoch. 

In the synthetic classification task, we split the datasets into three parts, i.e., the training, validation, and test datasets, following the convention of machine learning \cite{goodfellow2016deep}. The training dataset $\mathcal{D}_{tr}$ is used to optimize the trainable parameters  during the learning process. The function of the validation dataset $\mathcal{D}_{va}$ is estimating how well the  classifier has been trained. During $T$ epochs,  the trainable parameters that achieve  the highest validation accuracy are set as the output parameters. Mathematically, the output parameters satisfy
\begin{equation}
	\hat{\bm{\theta}} = \max_{\{\bm{\theta}^{(t)}\}_{t=1}^T} \sum_{i} \mathbbm{1}_{\tilde{y}^{(i)}(\bm{\theta}^{(t)}, \bm{x}^{(i)}) = y{(i)}},  
\end{equation}
where $\{\bm{x}^{(i)}, y^{(i)}\}\in \mathcal{D}_{va}$, $\tilde{y}^{(i)}$ is the prediction of the classifier given $\bm{\theta}^{(t)}$ and $\bm{x}^{(i)}$, and $\mathbbm{1}_{z}$ is the indicator function that takes the value $1$ if the condition $z$ is satisfied and zero otherwise. 
  Finally, the output parameters $\hat{\bm{\theta}}$ are applied to  the test dataset to benchmark the performance of the trained classifier.

\subsection{Implementation of the synthetic dataset}\label{append:ML_sub1}
Here we recap the method to construct  the  synthetic dataset proposed in \cite{havlivcek2019supervised}. Denote the encoding layer as
\begin{equation}
	U_{\bm{x}} = \RY(\bm{x}_1)\otimes \RY(\bm{x}_2)\otimes \RY(\bm{x}_3).
\end{equation}

To establish the synthetic dataset $\mathcal{D}$ used in the main text, we first generate a set of data points $\{\bm{x}^{(i)}\}$ with $\bm{x}^{(i)}\in\mathbb{R}^3$.  We then define the optimal circuit as 
\begin{equation}
U^*(\bm{\theta}^*) = \prod_{l=1}^3 U_l^*(\bm{\theta}_l^{*}),
\end{equation}  
where $U_l^*(\bm{\theta}_l^{*})=\otimes_{j=1}^3 \RY(\bm{\theta}_{l,j}^*) (\text{CNOT}\otimes I_2)(I_2\otimes \text{CNOT})$ and the parameter $\bm{\theta}^*_{l,j}$ is uniformly sampled from $[0, 2\pi)$ for all $j\in[3]$ and $l\in[3]$. The strategy to label $\bm{x}^{(i)}$ is  as follows. Let $\Pi=\mathbb{I}_4\otimes \ket{0}\bra{0}$ be the measurement operator. The data point $\bm{x}^{(i)}$ is labeled as  $y^{(i)}=1$ if 
\begin{equation}\label{eqn:append_ML_data_prep1}
\langle 000|U_{\bm{x}^{(i)}}^{\dagger} U^*(\bm{\theta}^*)^{\dagger} \Pi U^*(\bm{\theta}^*) U_{\bm{x}^{(i)}}|000 \rangle \geq 0.75.	
\end{equation}
The label of $\bm{x}^{(i)}$ is assigned as $y^{(i)}=0$ if 
\begin{equation}
\langle 000|U_{\bm{x}^{(i)}}^{\dagger} U^*(\bm{\theta}^*) ^{\dagger} \Pi U^*(\bm{\theta}^*) U_{\bm{x}^{(i)}}|000 \rangle \leq 0.25.	
\end{equation}
Note that, if the measured result is in the range $(0.25, 0.75)$, we drop this data point and sample a new one.  By repeating the above procedure, we can built the synthetic dataset $\mathcal{D}$.  

\subsection{Simulation results of the synthetic dataset classification and the fierce competition phenomenon}\label{append:ML_sub2}
Here we first introduce how to use the quantum kernel classifier to conduct the prediction. Given the data point $\bm{x}^{(i)}\in \mathcal{D}$ at the $t$-th epoch, the quantum kernel classifier is composed of two unitraies, i.e., $U_{\bm{x}^{(i)}}$ and $U(\bm{\theta}^{(t)})$, where the sequence of quantum gates in $U(\bm{\theta}^{(t)})$ is fixed as shown in Figure \ref{fig:ML_res} (b). The output of quantum kernel classifier yields
\begin{equation}
\tilde{y}(\bm{x}^{(i)},\bm{\theta}^{(t)}) = 	\langle 000|U_{\bm{x}^{(i)}}^{\dagger} U(\bm{\theta}^{(t)}) \Pi U(\bm{\theta}^{(t)}) U_{\bm{x}^{(i)}}|000 \rangle.
\end{equation}
The predicted label of $\bm{x}^{(i)}$, i.e., $g(\tilde{y}(\bm{x}^{(i)},\bm{\theta}^{(t)}))$, becomes
\begin{equation}
	g(\tilde{y}(\bm{x}^{(i)},\bm{\theta}^{(t)})) 
	=
	\begin{cases}
   0,& \text{if } \tilde{y}(\bm{x}^{(i)},\bm{\theta}^{(t)})< 0.5\\
    1,              & \text{otherwise}
\end{cases}. 
\end{equation}

When QAS is employed to enhance the trainability and to mitigate error of the quantum kernel classifier,  the arrangement of quantum gates in  $U(\bm{\theta})$ is no longer fixed and depends on the sampled ansatz. In other words, at the $t$-th epoch, given the data point $\bm{x}^{(i)}\in \mathcal{D}$, the measured result $\tilde{y}(\mathcal{A},\bm{x}^{(i)},\bm{\theta}^{(t)})$  in Eqn.~(\ref{eqn:loss_ML}) is  
\begin{equation}
\tilde{y}(\mathcal{A},\bm{x}^{(i)},\bm{\theta}^{(t)}) = 	\langle 000|U_{\bm{x}^{(i)}}^{\dagger} U(\bm{\theta}^{(t)}, \bm{a})^{\dagger} \Pi U(\bm{\theta}^{(t)}, \bm{a}) U_{\bm{x}^{(i)}}|000 \rangle,
\end{equation}
where $U(\bm{\theta}^{(t)}, \bm{a})$ denotes that the trainable unitary amounts to the ansatz $\bm{a}$ and the corresponding trainable parameters $\bm{\theta}^{(t)}$ are controlled by the supernet $\mathcal{A}$. 

\begin{figure*}[ht!]
	\centering
\includegraphics[width=0.99\textwidth]{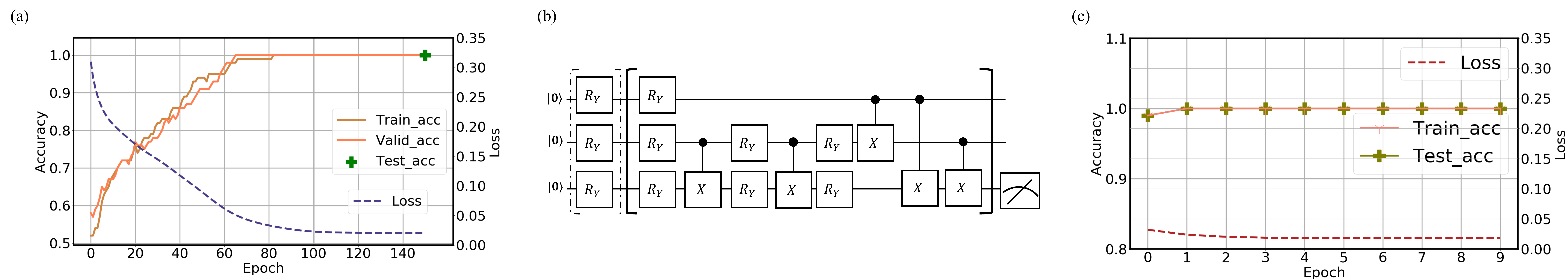} 
\caption{\small{\textbf{Simulation results for the synthetic data classification.} (a) The performance  of the conventional quantum classifier under the noiseless setting. The label `Train\_acc', `Valid\_acc', and `Test\_acc' refers to the  training, validation, and test accuracy, respectively. (b) The simulation results of QAS in the retraining stage under the noiseless setting. (c) The ansatz (i.e., the quantum circuit architecture) generated  by QAS in Step 3.  }}
\label{fig:supp_ML}
\end{figure*}

We then provide the simulation results of the conventional quantum kernel classifier and QAS towards the synthetic dataset $\mathcal{D}$ under the noiseless setting. As exhibited in Figure \ref{fig:supp_ML} (a),  both the training and validation accuracies of the conventional quantum kernel classifier fast converge to $100\%$ after $80$ epochs. The test accuracy also reaches $100 \%$, highlighted by the green marker.  Meanwhile,  the loss $\mathcal{L}$ decreases to $0.24$. These results indicate that the conventional quantum kernel classifier with the protocol as depicted in Figure \ref{fig:ML_res} (b) can well learn the synthetic dataset $\mathcal{D}$. 

The hyper-parameters of QAS under the noiseless setting are identical to the noisy setting introduced in the main text. Specifically, we set $T=400$ and $W=1$ in the training stage (Step 2), $K=500$ in the ranking stage (Step 3), and $T=10$ in the retraining stage (Step 4).  Figure \ref{fig:supp_ML} (b) demonstrates the output ansatz in Step 3. Compared to the conventional quantum kernel classifier, the output ansatz includes fewer CNOT gates, which is more amiable for physical implementations. Figure \ref{fig:supp_ML} (c)  illustrates the learning performance of the output ansatz in the retraining stage. Concretely, both the training and test accuracies converge to $100\%$ after one epoch. These results indicate that QAS can well learn the synthetic dataset $\mathcal{D}$ under the noiseless setting.  Note that for all simulation results related to classification tasks,   the Adam optimizer \cite{goodfellow2016deep} is exploited to update the training parameters of the quantum kernel classifier and QAS. The learning rate is set as $0.05$.   
    
We end this subsection by explaining the fierce competition phenomenon encountered in the optimization of QAS. Namely, when the number of supernets is $1$, some ansatze that can achieve high classification accuracies with independently training, will perform poorly in QAS. To exhibit that QAS indeed searches a set of ansatze (quantum circuit architectures) with high classification accuracies, we examine the correlation of the performance of the ansatz with independently optimization and training by QAS. In particular, we randomly sample $500$ ansatze from all possible architectures and evaluate the widely-used Spearman and Kendall tau rank correlation coefficients \cite{pirie2004spearman,kendall1938measure}, which are in the range of $[0, 1]$. 
In particular, larger correlation coefficients (or equivalently, stronger correlations) indicate that the ranking distribution achieved by QAS is consistent with the performance of different circuit architectures with independently training. Moreover, larger correlation coefficients also imply that the output ansatz of QAS can well estimate the target ansatz $\bm{a}^*$ in Eqn.~(\ref{eqn:obj_QAS}).

 The Spearman rank correlation coefficient $\rho_S$ quantifies the monotonic relationships between random variables $r$ and $s$. Specifically, the spearman rank correlation coefficient between $r$ and $s$ is defined as
 \begin{equation}
 	\rho_S=\frac{cov(r,s)}{\sigma_r\sigma_s},
 \end{equation}
where $cov(\cdot,\cdot)$ is the covariance of two variables, and $\sigma_r$ ($\sigma_s$) refers to the standard deviations of $r $ $(s)$. Suppose that $\bm{r}\in\mathbb{R}^n$ and $\bm{s}\in\mathbb{R}^n$ are two observation vectors of $r$ and $s$, respectively, the explicit form $\rho_S$ is
\begin{equation}
	\rho_S= 1 - \frac{6\sum_{i=1}^n(\bm{r}_i-\bm{s}_i)^2}{n(n^2 - 1)}.
\end{equation}
When the Spearman rank correlation is employed in QAS, the observation vector $\bm{r}$ ($\bm{s}$) corresponds to the achieved validation accuracy of the sampled $500$ ansatze in the ranking stage, while the observation vector $\bm{s}$ corresponds to the achieved validation accuracy of the sampled $500$ ansatze with independently training. 

The Kendall tau rank correlation coefficient concerns the relative difference of concordant pairs and discordant pairs. Specifically, in QAS, denote $\bm{r}$ ($\bm{s}$) as the observation vector that refers to the achieved validation accuracy of the sampled $500$ ansatze in the ranking stage (with independently training). Given any pair $(\bm{r}_i,\bm{r}_j)$ and $(\bm{s}_i,\bm{s}_j)$, it is said to be concordant if $(\bm{r}_i>\bm{r}_j)\land (\bm{s}_i>\bm{s}_j)$ or $(\bm{r}_i<\bm{r}_j)\land (\bm{s}_i<\bm{s}_j)$; otherwise, it is  disconcordant. According to the above definition, the explicit form of the  Kendall tau rank correlation coefficient is
\begin{equation}
	\rho_K= \frac{2}{n(n-1)}\sum_{i<j}\text{sign}(\bm{r}_i-\bm{r}_j) \text{sign}(\bm{s}_i-\bm{s}_j),
\end{equation} 
where $\text{sign}(\cdot)$ represents the sign function.

Table \ref{tab:correlation} summarizes the correlation coefficients with  $n=500$. Specifically, when the number of supernets is $1$, we have $\rho_K=0.113$, which implies that the correlation between $\bm{r}$ and $\bm{s}$ is very low. By contrast, with increasing the number of supernets to $5$ and $10$, the correlation coefficients   $\rho_S$ and $\rho_K$ are dramatically enhanced, which are $0.723$ and $0.536$, respectively. Moreover, when the number of supernets is $W=10$ and the number of iterations is increased to $T=1000$, the correlation coefficients   $\rho_S$ and $\rho_K$ can be further improved, which are $0.774$ and $0.591$, respectively. These results indicate that the competition phenomenon in QAS can be alleviated by introducing more supernets and increasing the number of training iterations. In doing so, the performance of ansatze evaluated by QAS can well accord with their real performance with independently training.   
  
  \begin{table}[h!]
\begin{tabular}{|l|l|l|l|l|}
\hline
         & $W=1$ \& $T=500$ & $W=5$ \& $T=500$ &   $W=10$ \& $T=500$ & $W=10$ \& $T=1000$\\ \hline
$\rho_S$ & 0.488            & 0.723 & 0.716    & 0.774    \\ \hline
$\rho_K$ & 0.113            & 0.536         & 0.531 & 0.591\\ \hline
\end{tabular}
\caption{\small{\textbf{The correlation coefficients.} The label `$W=a$ \& $T=b$' represents that the number of supernets and training iterations  is  $a$ and $b$, respectively.}}
\label{tab:correlation}
\end{table}

\subsection{The performance of QAS towards the noise model extracted from the real quantum devices}\label{appendix:subsec:QAS-ibmq-classifier}
\begin{figure}[ht]
\begin{minipage}[b]{0.3 \linewidth}
\centering
\includegraphics[width=0.65\textwidth]{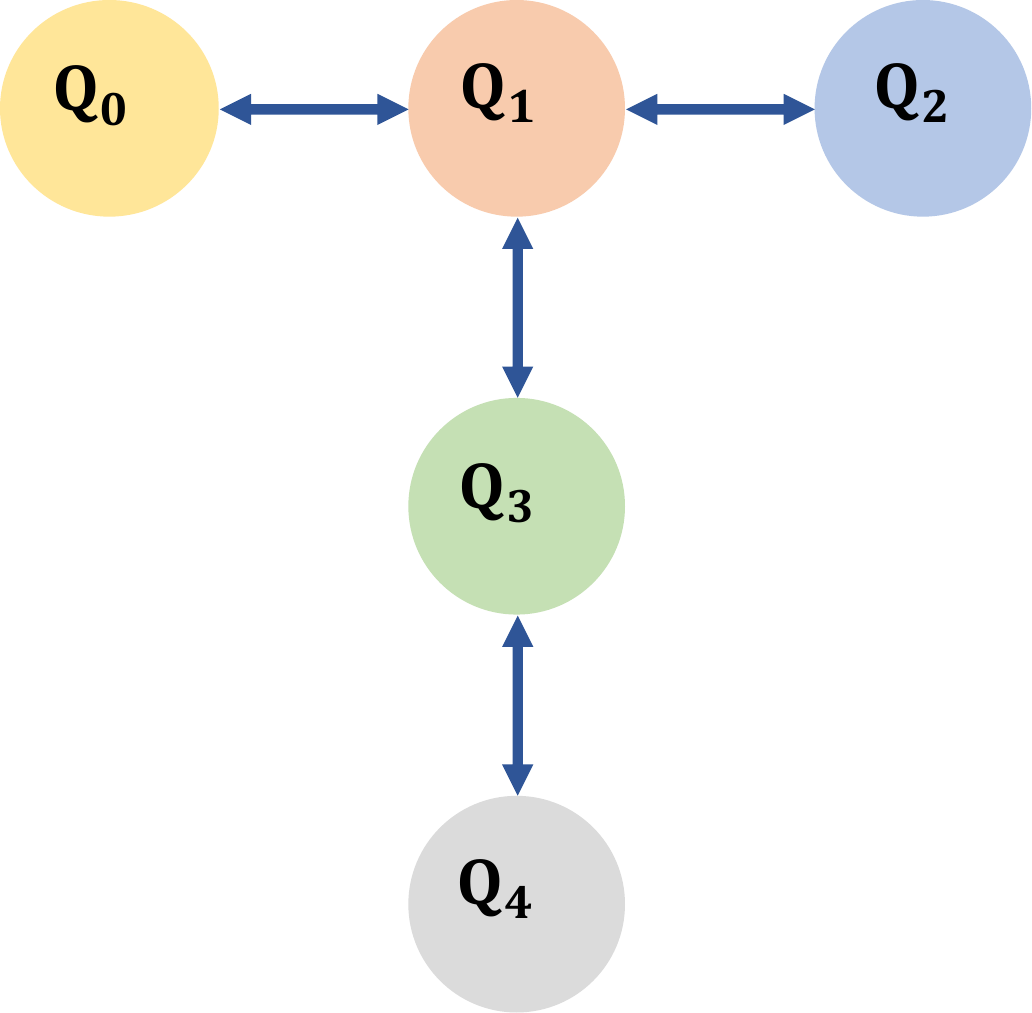}
\caption{\small{\textbf{The qubits connectivity of `Ibmq\_lima'.}}}
\label{fig:supp_ibm_qclass}
\end{minipage}
\hspace{0.5cm}
\begin{minipage}[b]{0.65\linewidth}
\centering
\begin{tabular}{|c|c|c|c|c|c|}
\hline
Qubit & T1($\mu s$) & T2($\mu s$ ) & Readout error & Single-qubit U2 error gate & CNOT error rate                                                                          \\ \hline
Q0    & 94.36       & 170.91       & 1.90E-2       & 1.87E-4                    & cx0-1: 6.76E-3                                                                           \\ \hline
Q1    & 108.32       & 140.36       & 2.38E-2       & 3.01E-4                    & \begin{tabular}[c]{@{}l@{}}cx1-0: 6.76E-3\\ cx1-2: 5.53E-3\\ cx1-3: 1.49E-2\end{tabular} \\ \hline
Q2    & 146.07      & 154.00      & 1.75E-2       & 1.92E-4                    & cx2-1: 5.53E-3                                                                           \\ \hline
Q3    & 106.5      & 89.98       & 3.44E-2       & 3.44E-4                    & \begin{tabular}[c]{@{}l@{}}cx3-1: 1.49E-2\\ cx3-4: 1.63E-2\end{tabular}                  \\ \hline
Q4    & 22.34       & 19.02      & 5.48E-2       & 6.77E-4                    & cx4-3: 1.63E-2                                                                           \\ \hline
\end{tabular}
\caption{\small{\textbf{Performance of qubits for  Ibmq\_lima.} T1 and T2 refer to the energy relaxation time and dephasing time, respectively. U2 and  CNOT gates error obtained via performing randomized benchmarking. The label `cxa-b' represents that the CNOT gate is applied to the qubits $a$ and $b$.  }}
\label{tab:my-table-lima}
\end{minipage}
\end{figure}

We evaluate the classification accuracy of the quantum classifier equipped with the hardware-efficient ansatz and the ansatz searched by QAS under the noise model extracted from a real quantum device, i.e., an IBM’s $5$-qubit quantum machine named `Ibmq\_lima'. The qubit connectivity of the deployed quantum machine is illustrated in Figure \ref{fig:supp_ibm_qclass} and its system parameters are summarized in Figure \ref{tab:my-table-lima}.  

The implementation details are as follows. The construction of the synthetic dataset is identical to those introduced in Supplementary \ref{append:ML_sub1}, except for setting the feature dimension as $5$ instead of $3$. For the quantum classifier with the hardware-efficient ansatz, the number of layers and the  number of epochs are set as $L=3$ and $T=400$, respectively.  The  hardware-efficient ansatz used in the baseline experiment takes the form $U(\bm{\theta}) = \prod_{l=1}^3 U_l(\bm{\theta}_l)$,  where $U_l(\bm{\theta}_l)=\otimes_{j=1}^5 \RY(\bm{\theta}_{l,j}) (\text{CNOT}\otimes \text{CNOT} \otimes I_2)(I_2\otimes \text{CNOT})\otimes \text{CNOT})$.   For QAS, the choices of single-qubit gates and two-qubit gates are $\{\RY,\RZ\}$ and $\{\CNOT, \mathbb{I}_4\}$, respectively. The $\CNOT$ gates can conditionally interact with four qubit-pairs, i.e., \{0-1, 1-2, 1-3, 3-4\}, following the topology of Ibmq\_lima. We apply two settings to evaluate the learning performance of QAS. In the first setting, we set the  number of supernets  as $W=1$ and the number of epochs in the optimization stage as $T=10$.  In the second setting, we set the  number of supernets as $W=5$ and the number of epochs   as $T=400$. For both settings, the number of layers is set as $L=3$ and the number of the sampled ansatze at the ranking stage is $K=500$.

\begin{figure}[h]
\centering
\includegraphics[width=0.99\textwidth]{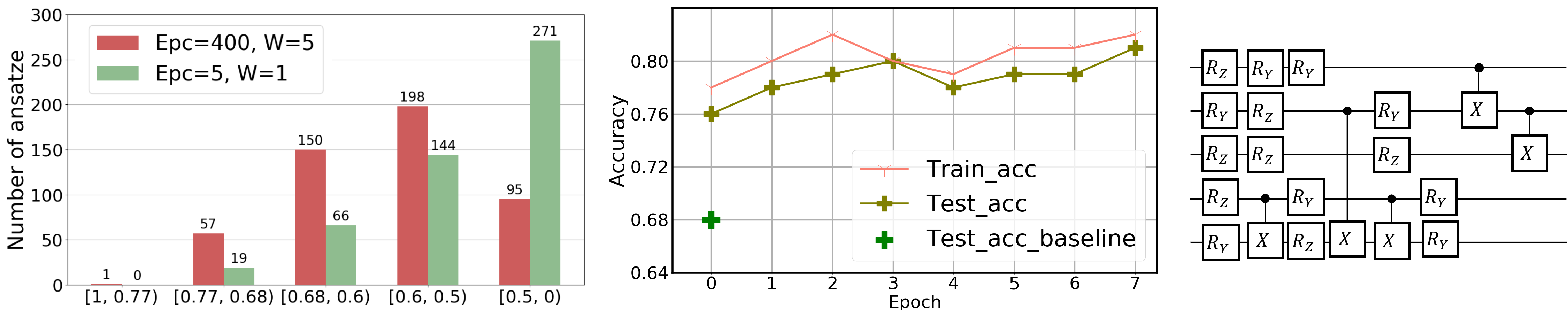}
\caption{\small{\textbf{Simulation results of the quantum classifiers under the noise model extracted from Ibmq\_lima}. The left panel illustrates the performance of QAS with two different settings after the optimization process.  The label `Epc=a, W =b' represents that the number of epochs and supernets is $T=a$ and $W=b$, respectively. The x-axis means that the validation accuracy of the sampled ansatz is in the range of $[c,d)$. The middle  panel exhibits the performance of the quantum classifiers with the hardware-efficient ansatz (labeled by `Test\_acc\_baseline')  and the ansatz searched by QAS (labeled by `Train/Test\_acc')  at the fine tuning stage under the noisy setting. The right panel depicts the searched ansatz before quantum compiling.}}	
\label{fig:ibmq-clc}
\end{figure}

The simulation results are exhibited in Figure \ref{fig:ibmq-clc}. For the quantum classifier with the hardware-efficient ansatz, the achieved test accuracy is $68\%$. We utilize this test accuracy as the baseline to quantify the learning performance of QAS.  As shown in the left panel of Figure \ref{fig:ibmq-clc}, for the first setting (i.e., $T=5$ and $W=1$), there are in total $19$ ansatze out of $K=500$ ansatze achieving a higher accuracy beyond the baseline. When we increase the number of epochs and the  number of supernets  to $T=400$ and $W=5$ (i.e., corresponding to the second setting), there are in total $58$ ansatze out of $K=500$ ansatze surpassing the baseline. Meanwhile, the average performance over the sampled $K=500$ ansatze is better than the first setting. As shown in the middle panel of Figure \ref{fig:ibmq-clc}, when we retrain the searched ansatz in the second setting  (depicted in the right panel of Figure \ref{fig:ibmq-clc}) with $8$ epochs, the test accuracy improves to $81\%$. These observations validate the effectiveness of QAS of enhancing the learning performance of VQAs towards classification tasks. Moreover, increasing the  number of supernets  $W$ and the number of epochs $T$ contributes to improve the capability of QAS.

\section{Experimental Details of the ground state energy estimation}\label{append:q_chem}
In this section, we first briefly  recap the ground state energy estimation task in Supplementary \ref{append:sub:q_chem1}.  In Supplementary \ref{append:sub:q_chem2}, we compare the performance of QAS and conventional VQE towards the ground state energy estimation task when they are implemented on real quantum hardware.

\subsection{The ground state energy estimation}\label{append:sub:q_chem1}
A central application of VQAs is solving the electronic structure problem, i.e., finding the ground state energies of chemical systems described by Hamiltonians. Note that chemical Hamiltonians in the second quantized basis set approach can always be mapped to a linear combination of products of local Pauli operators \cite{mcardle2020quantum}.  In particular, the explicit form of the molecular hydrogen Hamiltonian $H_h$ in Eqn.~(\ref{eqn:ground_H2}) is  
\begin{eqnarray}\label{eqn:appendchem1}
	H_h =  && -0.042 + 0.178 (Z_0 + Z_1) -0.243 (Z_2 + Z_3) + 0.171 Z_0Z_1  + 0.123(Z_0 Z_2 + Z_1Z_3) + 0.168(Z_0 Z_3 + Z_1 Z_2) \nonumber\\
	&& + 0.176Z_2Z_3 +  0.045 (Y_0X_1X_2Y_3 - Y_0Y_1X_2X_3 - X_0 X_1 Y_2 Y_3 + X_0 Y_1 Y_2 X_3). 
\end{eqnarray}
The goal of the variational Eigen-solver (VQE) is generating a parameterized wave-function $\ket{\Psi(\bm{\theta})}$ to achieve
 \begin{equation}
\min_{\bm{\theta}}  	|\langle \Psi(\bm{\theta}) |H_h| \Psi(\bm{\theta}) \rangle - E_m|. 
 \end{equation}
 The linear property of $H_h$ in Eqn.~(\ref{eqn:appendchem1}) implies that the value $|\langle \Psi(\bm{\theta}) |H_h| \Psi(\bm{\theta}) \rangle$ can be obtained by iteratively  measuring $\ket{\Psi(\bm{\theta})}$ using  Pauli operators in $H_h$, e.g.,  such as $|\langle \Psi(\bm{\theta}) |\mathbb{I}_8\otimes Z_0| \Psi(\bm{\theta}) \rangle$ and $|\langle \Psi(\bm{\theta}) |X_0Y_1Y_2X_3| \Psi(\bm{\theta}) \rangle$. The lowest energy of $H_h$ equals to $E_m=-1.136 \Ha$, where `$\Ha$' is the abbreviation of Hartree, i.e., a unit of energy used in molecular orbital calculations with $1 \Ha = 627.5 \text{kcal/mol}$. The exact value of $E_m $ is acquired from a full configuration-interaction calculation \cite{mcardle2020quantum}. 
 
We note that the quantum natural gradient optimizer \cite{stokes2020quantum}, which can accelerate the convergence rate, is employed to optimize the trainable parameters for both VQE and QAS, where the learning rate is set as $0.2$.

\subsection{The performance of QAS on real quantum devices}\label{append:sub:q_chem2}

\begin{figure}[ht]
\begin{minipage}[b]{0.3 \linewidth}
\centering
\includegraphics[width=0.65\textwidth]{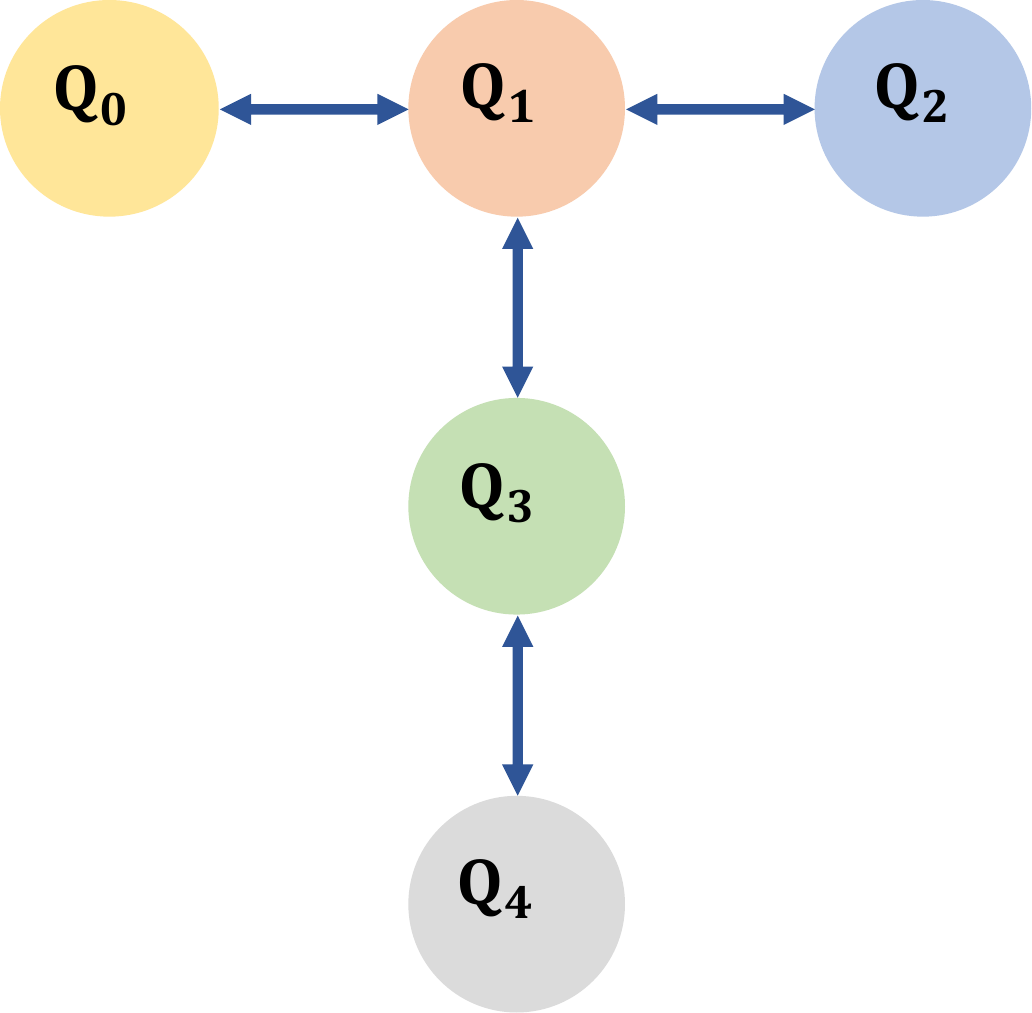}
\caption{\small{\textbf{The qubits connectivity of `Ibmq\_ourense'.}}}
\label{fig:supp_ibm_qchem}
\end{minipage}
\hspace{0.5cm}
\begin{minipage}[b]{0.65\linewidth}
\centering
\begin{tabular}{|c|c|c|c|c|c|}
\hline
Qubit & T1($\mu s$) & T2($\mu s$ ) & Readout error & Single-qubit U2 error gate & CNOT error rate                                                                          \\ \hline
Q0    & 75.75       & 50.81       & 1.65E-2       & 5.22E-4                    & cx0-1: 9.55E-3                                                                           \\ \hline
Q1    & 78.47       & 27.56       & 2.38E-2       & 4.14E-4                    & \begin{tabular}[c]{@{}l@{}}cx1-0: 9.55E-3\\ cx1-2: 9.44E-3\\ cx1-3: 1.25E-2\end{tabular} \\ \hline
Q2    & 101.51      & 107.00      & 1.57E-2       & 1.83E-4                    & cx2-1: 9.44E-3                                                                           \\ \hline
Q3    & 79.54       & 78.38       & 3.95E-2       & 4.30E-4                    & \begin{tabular}[c]{@{}l@{}}cx3-1: 1.25E-2\\ cx3-4: 8.34E-3\end{tabular}                  \\ \hline
Q4    & 74.27       & 30.00       & 4.74E-2       & 4.20E-4                    & cx4-3: 8.34E-3                                                                           \\ \hline
\end{tabular}
\caption{\small{\textbf{Performance of qubits for  Ibmq\_ourense.} T1 and T2 refer to the energy relaxation time and dephasing time, respectively. U2 and  CNOT gates error obtained via performing randomized benchmarking. The label `cxa-b' represents that the CNOT gate is applied to the qubits $a$ and $b$.  }}
\label{tab:my-table}
\end{minipage}
\end{figure}

Here we carry out QAS and the conventional VQE on IBM’s $5$-qubit quantum machine, i.e., `Ibmq\_ourense', to accomplish the ground state energy estimation of $H_h$. The qubit connectivity of `Ibmq\_ourense' is illustrated in Figure \ref{fig:supp_ibm_qchem}, and the system parameters of these five qubits are summarized in Figure \ref{tab:my-table}.

The implementation detail is as follows. The hyper-parameters of QAS are  $L=3$, $W=10$, $K=500$, and $T=500$. To examine the compatibility of QAS, we restrict its searching spaces to be consistent with the qubit connectivity of `IBM\_ourense', i.e., the single-qubit gates are sampled from $\RY$ and $\RZ$, and CNOT gates can conditionally apply to the qubits pair $(0,1)$, $(1,0)$, $(1,2)$, $(2,1)$, $(1,3)$, and $(3,1)$, based on Figure \ref{tab:my-table}. We call this setting as QAS with the real connectivity (QAS-RC). Under such a setting, the number of all possible circuit architectures for QAS-RC is $1024^3$. The hyper-parameters setting for VQE are $L=3$ and $T=500$. The heuristic circuit architecture used in VQE is identical to the case introduced in the main text (Figure \ref{fig:chem_res} (a)). In the training process, we optimize VQE and QAS on classical computers under a noisy environment provided by the Qiskit package, which can approximately simulate the quantum gates error and readout error in `Ibmq\_ourense'. The reason that we move the training stage on the classical numerical simulators is because training VQE and QAS on `Ibmq\_ourense' will take an unaffordable runtime, due to the fair share run mode \cite{Qiskit}.

 \begin{figure*}[h!]
 	\centering
\includegraphics[width=0.9\textwidth]{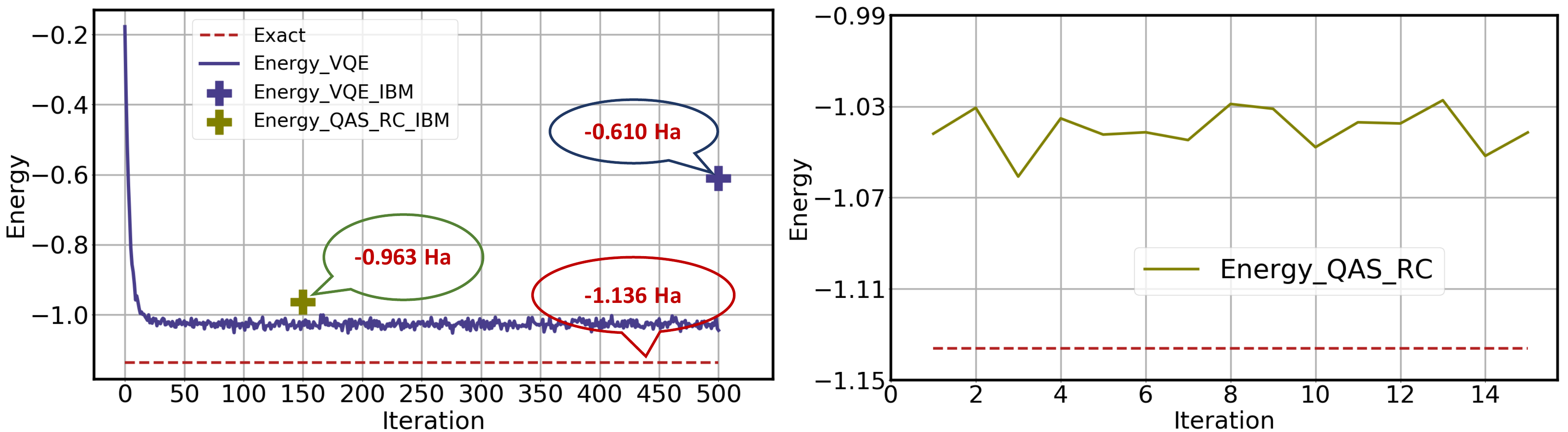}
\caption{\small{\textbf{Experiment results of the ground state energy estimation of VQE and  QAS.} In the left panel, the labels `Exact' and `Energy\_VQE' correspond to the exact ground state energy and the estimated energy of VQE obtained in the training process (achieved by  numerical simulators), respectively. In the left panel, the label  `Energy\_QAS\_RC' refer to the estimated energy of QAS-RC in the retraining phase.  The estimated energy of VQE and QAS-RC when they are implemented on the real quantum processor `Ibmq\_ourense' are indicated by two cross  markers, where the corresponding labels are  `Energy\_VQE\_IBM' and `Energy\_QAS\_RC\_IBM', respectively.  }}
\label{fig:supp_ibm_qchem1}
 \end{figure*}  
   
 The training performance of VQE and QAS-RC is demonstrated in  Figure \ref{fig:supp_ibm_qchem1}. In  particular, as shown in the left panel, the estimated ground energy by VQE is around  $-1.02 \Ha$ after $30$ iterations, highlighted by the dark blue line. The  performance of QAS-RC is shown in the right panel. Concretely, when we retrain the output ansatz with $15$ iterations, its estimated energy slightly oscillates around  $-1.04\Ha$, highlighted by the green solid line.  When we implement the optimized VQE and the optimized output ansatz of QAS-RC on the real quantum device, i.e.,  `Ibmq\_ourense',  their performances are varied. Specifically,  as demonstrated in the left panel of Figure \ref{fig:supp_ibm_qchem1},  the estimated ground energy by VQE is $-0.61\Ha$ (highlighted by the blue marker), while the estimated ground energy by QAS-RC is  $-0.963\Ha$ (highlighted by the green marker). Compared with VQE, the estimated result of QAS-RC  is much closer to the exact result. We utilize the following formula to quantify the relative deviation between the simulation and experiment results. Denoted the estimated energy obtained by the numerical simulation as $E_s$ and the test energy achieved by `Ibmq\_ourense' as $E_t$, the relative deviation follows
 
 \begin{equation}
 	err = \frac{|E_s - E_t|}{E_m},
 \end{equation}   
 where $E_m=-1.136$ is the exact result. Following this formula, the relative deviation for VQE and QAS-RC is $36.1\%$ and $6.8\%$, respectively. Compared with the heuristic circuit architecture used in VQE, QAS that concerns the real qubits connectivities can dramatically reduce the relative deviation.  The above results not only indicate the compatibility of QAS, but also demonstrate that QAS can well adapt to the weighted gates noise and achieve a high performance towards quantum chemistry tasks.  
   
 \begin{figure*}[h!]  
 	\centering
 \includegraphics[width=0.999 \textwidth]{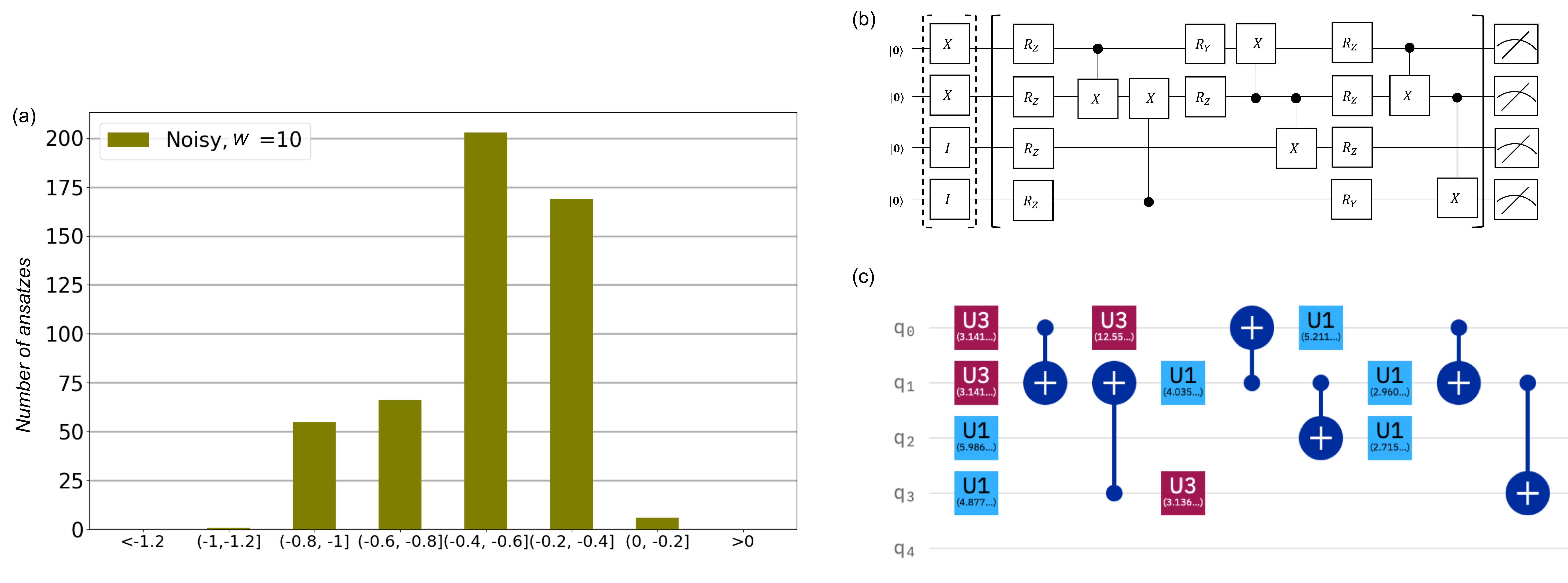}
 \caption{\small{\textbf{The simulation results of QAS-RC in the ranking stage.} The panel (a) demonstrates the  ansatze ranking distributions with $K=500$ of QAS-RC. The x-axis refers to the estimated energy of the given ansatz is in the range of $(a\Ha, b\Ha]$ with $a,b\in\mathbb{R}$. For most ansatze, their estimated energies are above $-0.2\Ha$.  The middle panel (b) exhibits the output ansatz of QAS-RC. The right panel (c) shows the implementation of the output ansatz of QAS-RC  on `Ibmq\_ourense'. }}
 \label{fig:app_q_chem_2}
 \end{figure*}

Finally, we compare the output ansatz of QAS-RC with the heuristic circuit architecture used in VQE.  The simulation results of  QAS-RC in the ranking stage are  summarized in Figure \ref{fig:app_q_chem_2}. In particular, the left panel (a) exhibits the ranking distributions of QAS-RC, where the estimated ground energy of  most ansatze concentrates on $[-0.6 \Ha, -0.4\Ha ]$. Figure \ref{fig:app_q_chem_2} (b) shows the output ansatz of QAS-RC, where the corresponding circuit implementation on `IBM\_ourense' is exhibited in Figure \ref{fig:app_q_chem_2} (c).  Compared with the heuristic circuit architecture used in VQE   (Figure \ref{fig:chem_res} (a)), the output ansatz of  QAS-RC contains fewer CNOT gates. This implies that QAS-RC has the ability to appropriately reduce the number of two-qubit gates to avoid introducing too much error, while the expressive power of the trainable circuit $U(\bm{\theta})$ can be well preserved. In other words, QAS can adapt to the weighted gate noise to seek the best circuit architecture.

 \section{Improving the ranking stage of QAS}\label{append:evolution}
  Recall the ranking stage of QAS, i.e., Step 3 of Figure \ref{fig:QAS}, is uniformly sampling $K$ ansatze from the supernet $\mathcal{A}$. The aim of this step is sampling the one, among the sampled ansatz, with the best performance. However, the uniformly sampling method implies that the sampled ansatze maybe come from $\mathcal{S}_{\text{bad}}$ with a high probability when $|\mathcal{S}_{\text{bad}}|>|\mathcal{S}_{\text{good}}|$. It is highly desired to devise more effective sampling methods.

Here we utilize an evolutionary algorithm, i..e, nondominated sorting genetic algorithm II (NSGA-II) \cite{deb2002fast}, to facilitate the ansatze ranking problem.  The intuition behind employing NSGA-II is actively searching potential ansatze with good performance instead of uniformly sampling ansatze from all possible circuit architectures.  Note that several recent studies, e.g., Refs \cite{chivilikhin2020mog,rattew2019domain}, have directly utilized the evolutionary and multi-objective genetic algorithms to complete ansatz design.

\begin{figure}[h!]
	\centering
\includegraphics[width=0.5\textwidth]{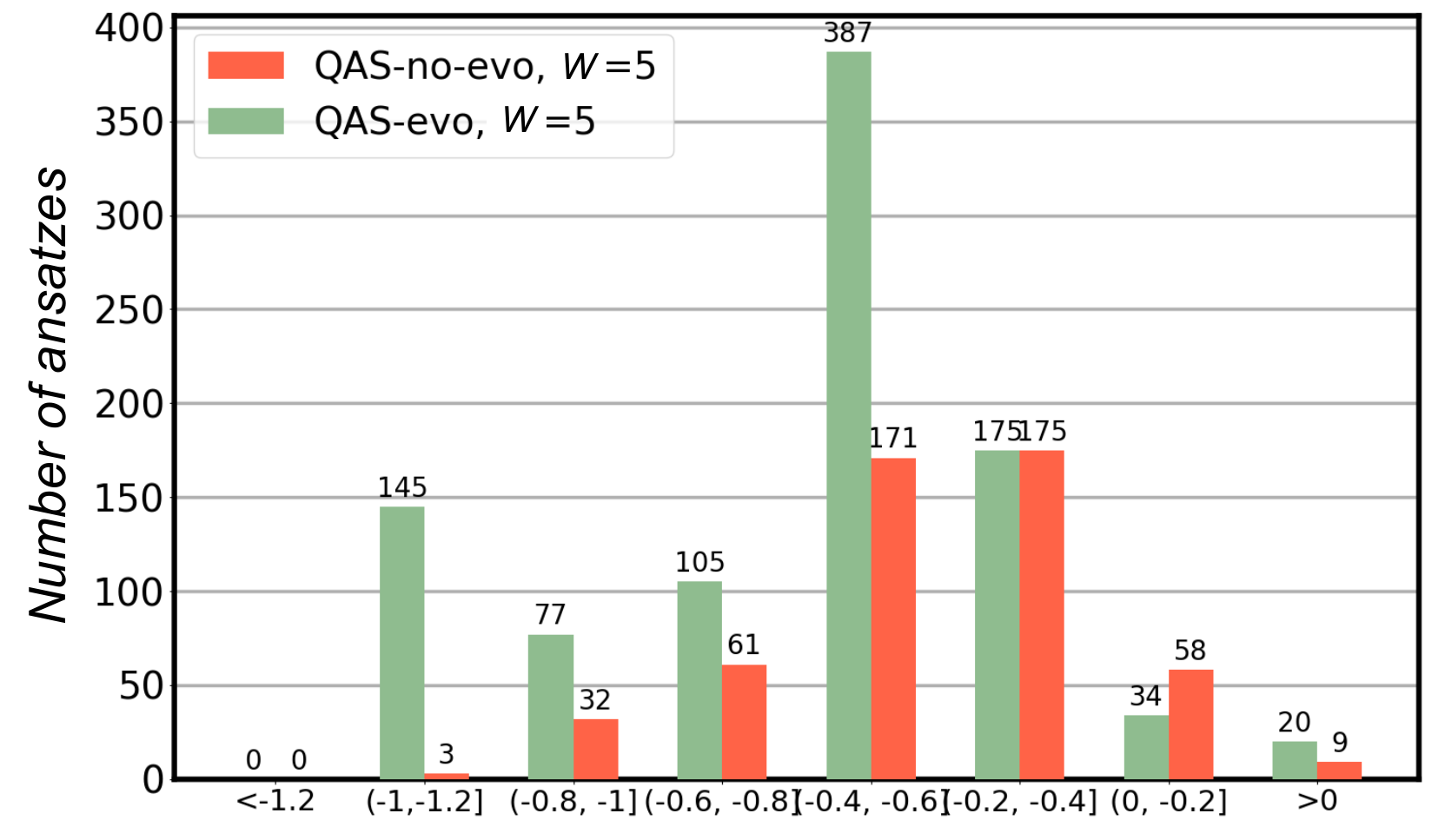}
\caption{\small{\textbf{Simulation result of QAS assisted by NSGA-II.} The label `QAS-no-evo, W=5' and `QAS-evo, W=5' refer to the QAS introduced in the main text and QAS assisted by evolutionary algorithm  with the number of supernets being $W=5$, respectively. The x-axis refers that	 the estimated energy of the given ansatz is in the range of $(a\Ha, b\Ha]$ with $a,b\in\mathbb{R}$. }}
\label{fig:supple_q_chem_evo}
\end{figure}

We apply QAS with the evolutionary algorithm  to tackle the ground state energy estimation problem described in the main text. Note that all hyper-parameters settings are identical to the uniformly sampling case, except for the settings related to the evolutionary algorithm. Particularly, we set the population size as $N_{pop}=50$ and the number of generations as $G_T=20$.  The simulation results under the noiseless setting are shown in Figure \ref{fig:supple_q_chem_evo}. In particular, QAS assisted by NSGA-II searches in total $943$ ansatze, and the estimated energy of $143$ ansatze ($15.2\%$) lies in the range from $-1\Ha$ to $-1.2\Ha$. By contrast, QAS with uniformly sampling strategy only finds $3$ ansatze among in total $500$ ansatze ($0.6\%$) in the same range. This result empirically confirms that evolutionary algorithms can advance the performance of QAS. 
 
We remark that other advanced machine learning techniques such as reinforcement learning \cite{sutton1998introduction} can also be exploited to benefit the performance of QAS.

\section{An empirical exploration for the trainability of QAS}\label{appendix:barren plateaus}
Here we empirically investigate the trainability of QAS through the lens of barren plateaus \cite{mcclean2018barren,marrero2020entanglement,patti2020entanglement,volkoff2021large}. Recall the main conclusion of the barren plateaus is that the gradient vanishes exponentially in the qubits count $N$. Mathematically, the expectation of the gradient norm of the objective function in Eqn.~(\ref{eqn:exp_erm}) tends to be zero and the corresponding variance will fast converge to zero with respect to  $N$, i.e., $\text{Var}_{\bm{\theta}}(\|\nabla_{\bm{\theta}} \mathcal{L}(\bm{\theta})\|)\sim O(e^{-LN})$.  With this regard, barren plateaus can be utilized as a measure to quantify the trainability of quantum algorithms. That is, when an algorithm  experiences a less impact of barren plateaus, it could possess a better trainability.

\begin{figure}[h!]
	\centering
\includegraphics[width=0.99\textwidth]{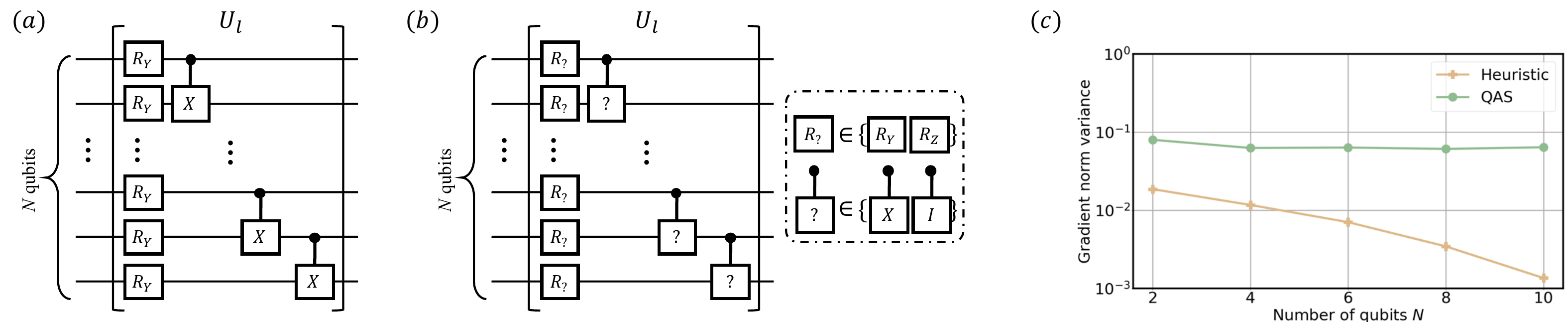}
\caption{\small{\textbf{Numerical simulations related to barren plateaus}. (a) The circuit implementation of the hardware-efficient ansatz. (b) The circuit implementation of the ansatze pool used in QAS.  (c) The variance of the gradient norm versus the number of qubits $N$. The label `heuristic' and `QAS' refers to the hardware-efficient ansatz and the ansatze pool used in QAS, respectively.  }}
\label{fig:barren}
\end{figure}

Following the above explanations, we conduct the following numerical simulations to demonstrate that the alleviation of barren plateaus in QAS. In particular, we compare the variance of the gradient norm, i.e., $\text{Var}_{\bm{\theta}}(\|\nabla_{\bm{\theta}} \mathcal{L}(\bm{\theta})\|)$, with respect to the hardware-efficient ansatz and the ansatze pool implied by QAS. The mathematical expression of the objective function is
\begin{equation}
\mathcal{L}=\Tr(HU(\bm{\theta})\rho U(\bm{\theta})^\dagger)~,	
\end{equation} 
where the observable $H$ equals to $\mathbb{I}_{2^{N-1}}\otimes\ket{0}\bra{0}$, the input state is $\rho=(\ket{0}\bra{0})^{\otimes N}$, and $U(\bm{\theta})$ corresponds to the hardware-efficient ansatz or the ansatz explored in QAS. For the hardware-efficient ansatz, we set the layer number as $L=3$, i.e., $U(\bm{\theta})=\prod_{l=1}^LU_l(\bm{\theta})$ and the implementation of $U_l(\bm{\theta})$ is shown in Figure \ref{fig:barren} (a). The calculation of $\text{Var}_{\bm{\theta}}(\|\nabla_{\bm{\theta}} \mathcal{L}(\bm{\theta})\|)$ is completed by randomly sampling $\bm{\theta}$ from a uniform distribution with $2000$ times. For QAS, the ansatze pool $\mathcal{S}$ is constructed by tailoring the hardware-efficient ansatz $U(\bm{\theta})$ introduced above. As shown in Figure \ref{fig:barren} (b), for each $U_l$ with $l\in [L]$, there are two choices of the single-qubit gates (i.e., $\RY$ and $\RZ$) and two choices of the two-qubit gates (i.e., $\CNOT$ and an identity operation). The calculation of $\text{Var}_{\bm{\theta}}(\|\nabla_{\bm{\theta}} \mathcal{L}(\bm{\theta})\|)$ is completed by  sampling $2000$ different ansatzes and sampling  one random $\bm{\theta}$ from a uniform distribution for each ansatz. The number of qubits $N$ ranges from $2$ to $10$.

The simulation results under the noiseless setting are shown in Figure \ref{fig:barren} (c). For the hardware-efficient ansatz, the variance of the gradient norm is continuously decreased with respect to the increased $N$.  This result can be  treated as an evidence of barren plateaus. By contrast, for the ansatze pool explored by QAS, the variance of the gradient norm for $N=4, 6, 8, 10$ is almost the same with each other. Meanwhile, for the same $N$, the variance of the gradient norm corresponding to the ansatze pool explored by QAS is always higher than that of the hardware-efficient ansatz.  Recall that Ref.~\cite{mcclean2018barren} states that the variance of gradients is continuously deceased with respect to the increased  $N$ and $L$, which induces the barren plateau phenomena for the sufficiently large $N$ and $L$. Nevertheless, according to the  simulation results in Figure \ref{fig:barren} (c), QAS does not obey such a tendency. These observations imply the potential of QAS to alleviate the influence of the barren plateaus.

\section{Progressive QAS for solving large-scale problems}\label{append:pro-QAS}
The proposed QAS introduced in the main text is only a prototype towards automatically seeking a good ansatz instead of the handcraft design. Namely, even though QAS utilizes weight sharing strategy to reduce the parameter space to $O(dLQ^N)$, there still exists an exponential dependence with $N$. This exponential relation implies that in certain cases, the searched ansatz by QAS may not well estimate the optimal ansatz $\bm{a}^*$ when $N$ becomes large within a reasonable runtime complexity. In this section, we devise a variant of QAS, termed as progressive QAS (Pro-QAS), to dramatically improve the learning performance of QAS for large-scale problems.

\subsection{Algorithmic implementation of Pro-QAS}
The key concept behind Pro-QAS is  narrowing the size of ansatze pool to ensure its performance. Different from QAS that directly samples an ansatz from $\mathcal{S}$ to conduct optimization, Pro-QAS seeks the targeted ansatz in a progressive way. Namely, given a hardware-efficient ansatz $U(\bm{\theta})$ in Eqn.~(\ref{eqn:UL_def}), Pro-QAS first freezes the gate arrangement of $U_{l}(\bm{\theta})$ with $l\neq 1$ and search the best gate arrangement of $U_{l=1}(\bm{\theta})$. Such a searching process is the same with Step 2 (optimization) and Step 3 (Ranking $K$ ansatze) in the original QAS. Once the search is completed, Pro-QAS begins to optimize the gate arrangement at the second layer $U_{l=2}(\bm{\theta})$ and freezes the rest $L-1$ layers. After progressively optimizing the gate arrangement of $L$ layers, the established ansatz $\bm{a}^{(T)}$ and its corresponding parameters $\bm{\theta}^{(T)}$ are used to approximate the optimal result $(\bm{a}^{*},\bm{\theta}^{*})$ in Eqn.~(\ref{eqn:exp_erm}). Notably, similar ideas of progressively constructing and optimizing ansatz have been exploited in Refs. \cite{grimsley2019adaptive,skolik2021layerwise,tang2021qubit,zhu2020adaptive}. Note that Ref.~\cite{campos2021abrupt} observed that an abrupt transition phenomenon for the progressive strategy. That is, when the cost function has the identity extrema and the number of layers is less than a critical value, the layer-wise training strategy could lead to an unfavorable performance. These results can be employed as guidance to improve the learning performance of Pro-QAS. For instance, the cost function adopted in Pro-QAS should be carefully designed to avoid the   identity extrema.

We then analyze the required runtime complexity and the reduced search space of Pro-QAS. Compared with the original QAS, the only difference of Pro-QAS is involving an extra outer loop to progressively optimize $L$ layers. Hence, in conjunction with the runtime complexity cost of QAS derived in Method, we conclude that the execution of Pro-QAS takes at most $O(dQNL)$ memory and $O(QNL^2)$ runtime. As for the size of search space, the progressively searching strategy decreases the size of the ansatze pool to $O(LQ^N)$, which is exponentially less than that of QAS in terms of $L$. Remarkably, such space can be further reduced when we progressively search the gate arrangement of each layer along the index of qubits. In this way, the search space of possible ansatze scales with $O(QNL)$, while the price to pay is linearly increasing the runtime cost with respect to $N$.

\subsection{Numerical simulation results of Pro-QAS}
We conduct numerical simulations to demonstrate the capability of the proposed Pro-QAS towards large-scale problems. In particular, we apply Pro-QAS to achieve a binary classification task. The construction rule of the dataset $\mathcal{D}=\{\bm{x}^{(i)}, y^{(i)}\}_{i=1}^{300}$ mainly follows Supplementary  \ref{append:ML_sub1}, where the only difference is enhancing the feature dimension of the input example from $3$ to $7$ and $10$, respectively. In other words, the number of qubits to load the input example $\bm{x}^{(i)}$ is $N=7$ (or $N=10$), which is remarkably larger than the classification task discussed in the main text with $N=3$. 

The hyper-parameters setting are as follows. For the case of $N=7$, the number of supernet is set as $W=1$ and the layer number is $L=5$.  The number of epochs in Step 2 is set as $T=200$. The allowed types of quantum gates are $\{\RY,\RZ, \CNOT \}$ with $Q=3$ and the qubits connectivity follows the chain structure. This setting implies that the total number of ansatze without any operation is $|\mathcal{S}| =2^{40}$. The depolarization channel is employed to simulate the quantum system noise. The depolarization rates for the single-qubit and two-qubit gates are set as $p=0.05$ and $p=0.1$, respectively. The number of sampled ansatze at the ranking stage is $K=128$.  For the case of $N = 10$, all settings are the same with the above one, except for setting $L=3$. To this end, the total number of ansatze is $|\mathcal{S}| = 2^{33}$.  

\begin{figure}[h!]
	\centering
	\includegraphics[width=0.98\textwidth]{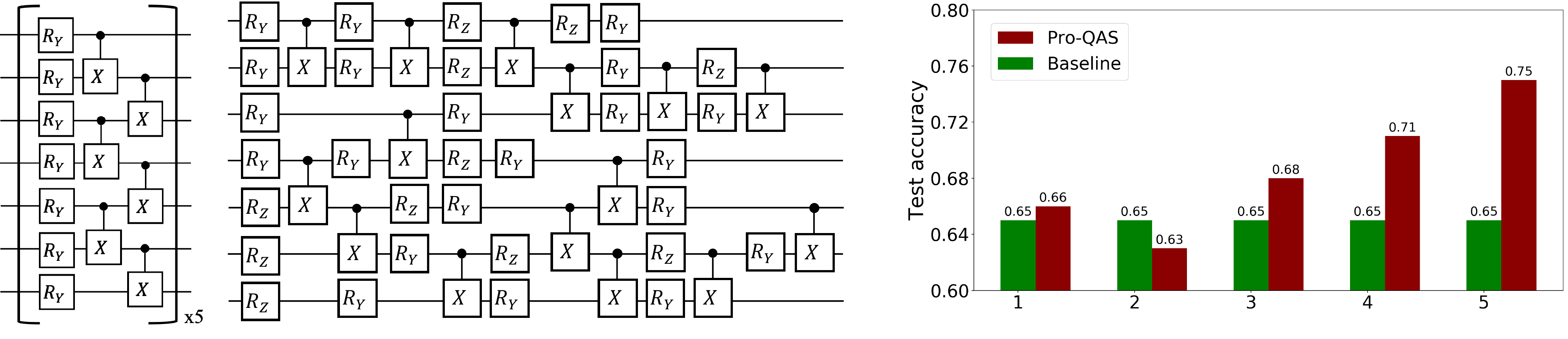}
	\caption{\small{\textbf{Simulation result for the $7$-qubit case.} The left subplot illustrates the implementation of the employed hardware-efficient ansatz. The middle subplot exhibits the ansatz searched by Pro-QAS. The right subplot compares the test accuracy between the hardware-efficient ansatz and the ansatz searched by Pro-QAS. The x-axis refers that the optimization of the $l$-th layer for Pro-QAS is accomplished. }}
	\label{fig:sim-res-7qubits}
\end{figure}

\begin{figure}[h!]
	\centering
	\includegraphics[width=0.98\textwidth]{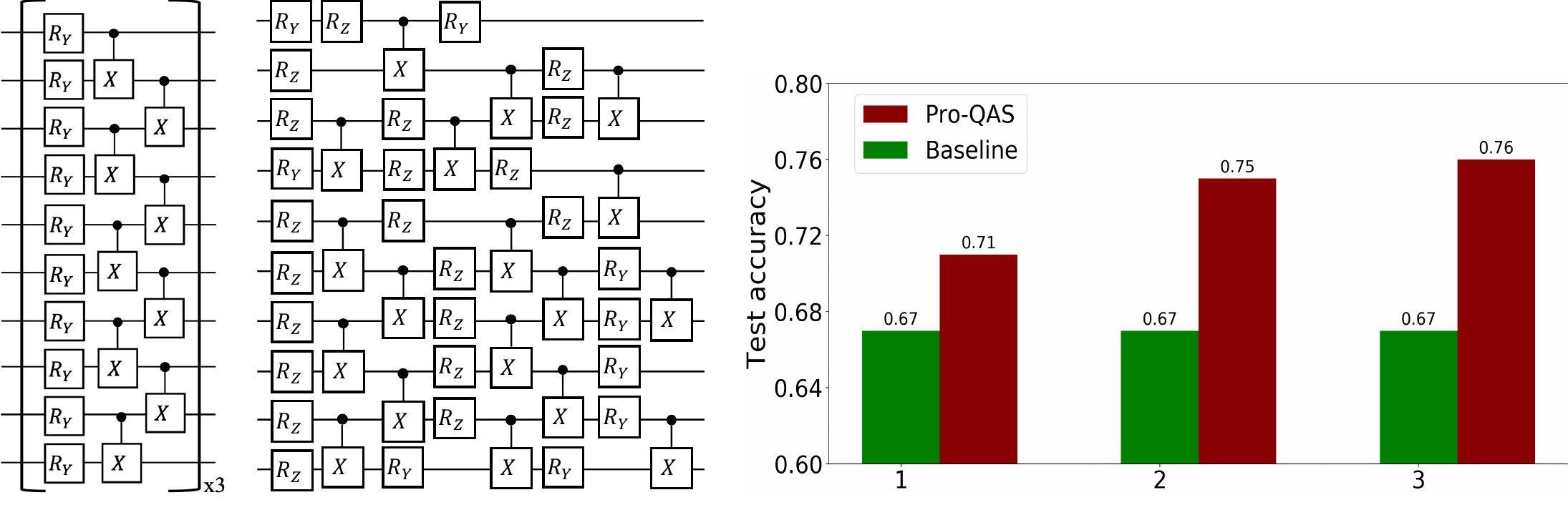}
	\caption{\small{\textbf{Simulation result for the $10$-qubit case.} The left subplot illustrates the implementation of the employed hardware-efficient ansatz. The middle subplot exhibits the ansatz searched by Pro-QAS. The right subplot compares the test accuracy between the hardware-efficient ansatz and the ansatz searched by Pro-QAS. The x-axis has the same meaning with the one introduced Figure \ref{fig:sim-res-7qubits}. }}
	\label{fig:sim-res-10qubits}
\end{figure}

The simulation results for the case of $N=7$ are exhibited in Figure \ref{fig:sim-res-7qubits}. As a reference, we employ the hardware-efficient ansatz with the identical layer number $L=5$, as shown in the left subplot, to learn the same dataset $\mathcal{D}$ \cite{havlivcek2019supervised}. After optimizing $200$ epochs, the test accuracy of the hardware-efficient ansatz converges to $65\%$, which is exploited as the baseline. The center subplot illustrates the ansatz searched by Pro-QAS. Compared with the hardware-efficient ansatz, the number of $\CNOT$ gates reduces from $30$ to $14$, which suppresses noise  by removing the unnecessary two-qubit gates. The achieved test accuracy is demonstrated in the right subplot. Specifically, when the optimization of the third layer is finished, Pro-QAS outperforms the baseline. When all $L$ layers are optimized, the ansatz searched by Pro-QAS attains $75\%$ test accuracy. The simulation results for $N=10$ are exhibited in Figure \ref{fig:sim-res-10qubits}.

The simulation results for the case of $N=10$ are exhibited in Figure \ref{fig:sim-res-10qubits}. Analogous to the case of $N=7$, the hardware-efficient ansatz with $L=3$ achieves $67\%$ test accuracy, which is employed as the baseline. The results shown in the right subplot evidence that the ansatz searched by Pro-QAS is superior to the hardware-efficient ansatz. Concretely, after searching, the test accuracy improves to $76\%$.  The center subplot illustrates the ansatz searched by Pro-QAS. To alleviate system noise and improve learning performance, the number of $\CNOT$ decreases from $27$ to $18$.    
  

\begin{thebibliography}{10}

\bibitem{cerezo2020variational2}
Marco Cerezo, Andrew Arrasmith, Ryan Babbush, Simon~C Benjamin, Suguru Endo,
  Keisuke Fujii,  et~al.
\newblock Variational quantum algorithms.
\newblock {\em Nat Rev Phys}, 3(9):625--644, 2021.

\bibitem{bharti2021noisy}
Kishor Bharti, Alba Cervera-Lierta, Thi~Ha Kyaw, Tobias Haug, Sumner
  Alperin-Lea, Abhinav Anand,  et~al.
\newblock Noisy intermediate-scale quantum algorithms.
\newblock {\em Rev. Mod. Phys.}, 94(1):015004, 2022.

\bibitem{beer2020training}
Kerstin Beer, Dmytro Bondarenko, Terry Farrelly, Tobias~J Osborne, Robert
  Salzmann, Daniel Scheiermann, et~al.
\newblock Training deep quantum neural networks.
\newblock {\em Nat Commun}, 11(1):1--6, 2020.

\bibitem{farhi2018classification}
Edward Farhi and Hartmut Neven.
\newblock Classification with quantum neural networks on near term processors.
\newblock {\em arXiv preprint arXiv:1802.06002}, 2018.

\bibitem{schuld2019quantum}
Maria Schuld and Nathan Killoran.
\newblock Quantum machine learning in feature hilbert spaces.
\newblock {\em Phys. Rev. Lett.}, 122(4):040504, 2019.

\bibitem{peruzzo2014variational}
Alberto Peruzzo, Jarrod McClean, Peter Shadbolt, Man-Hong Yung, Xiao-Qi Zhou,
  Peter~J Love, et~al.
\newblock A variational eigenvalue solver on a photonic quantum processor.
\newblock {\em Nat Commun}, 5:4213, 2014.

\bibitem{wang2019accelerated}
Daochen Wang, Oscar Higgott, and Stephen Brierley.
\newblock Accelerated variational quantum eigensolver.
\newblock {\em Phys. Rev. Lett.}, 122(14):140504, 2019.

\bibitem{stokes2020quantum}
James Stokes, Josh Izaac, Nathan Killoran, and Giuseppe Carleo.
\newblock Quantum natural gradient.
\newblock {\em Quantum}, 4:269, 2020.

\bibitem{mitarai2019generalization}
Kosuke Mitarai, Tennin Yan, and Keisuke Fujii.
\newblock Generalization of the output of a variational quantum eigensolver by
  parameter interpolation with a low-depth ansatz.
\newblock {\em Phys. Rev. Applied}, 11(4):044087, 2019.

\bibitem{preskill2018quantum}
John Preskill.
\newblock Quantum computing in the nisq era and beyond.
\newblock {\em Quantum}, 2:79, 2018.

\bibitem{havlivcek2019supervised}
Vojt{\v{e}}ch Havl{\'\i}{\v{c}}ek, Antonio~D C{\'o}rcoles, Kristan Temme,
  Aram~W Harrow, Abhinav Kandala, Jerry~M Chow, et~al.
\newblock Supervised learning with quantum-enhanced feature spaces.
\newblock {\em Nature}, 567(7747):209, 2019.

\bibitem{huang2021experimental}
He-Liang Huang, Yuxuan Du, Ming Gong, Youwei Zhao, Yulin Wu, Chaoyue Wang, et~al.
\newblock Experimental quantum generative adversarial networks for image
  generation.
\newblock {\em Phys. Rev. Applied}, 16(2):024051, 2021.

\bibitem{kandala2017hardware}
Abhinav Kandala, Antonio Mezzacapo, Kristan Temme, Maika Takita, Markus Brink,
  Jerry~M Chow, et~al.
\newblock Hardware-efficient variational quantum eigensolver for small
  molecules and quantum magnets.
\newblock {\em Nature}, 549(7671):242--246, 2017.

\bibitem{google2020hartree}
Google~AI Quantum et~al.
\newblock Hartree-fock on a superconducting qubit quantum computer.
\newblock {\em Science}, 369(6507):1084--1089, 2020.

\bibitem{holmes2021connecting}
Zo{\"e} Holmes, Kunal Sharma, Marco Cerezo, and Patrick~J Coles.
\newblock Connecting ansatz expressibility to gradient magnitudes and barren
  plateaus.
\newblock {\em PRX Quantum}, 3(1):010313, 2022.

\bibitem{benedetti2019parameterized}
Marcello Benedetti, Erika Lloyd, Stefan Sack, and Mattia Fiorentini.
\newblock Parameterized quantum circuits as machine learning models.
\newblock {\em Quantum Sci. Technol.}, 4(4):043001, 2019.

\bibitem{caro2021generalization}
Matthias~C Caro, Hsin-Yuan Huang, M~Cerezo, Kunal Sharma, Andrew Sornborger,
  Lukasz Cincio, et~al.
\newblock Generalization in quantum machine learning from few training data.
\newblock {\em arXiv preprint arXiv:2111.05292}, 2021.

\bibitem{du2018expressive}
Yuxuan Du, Min-Hsiu Hsieh, Tongliang Liu, and Dacheng Tao.
\newblock Expressive power of parametrized quantum circuits.
\newblock {\em Phys. Rev. Research}, 2:033125, Jul 2020.

\bibitem{du2021efficient}
Yuxuan Du, Zhuozhuo Tu, Xiao Yuan, and Dacheng Tao.
\newblock Efficient measure for the expressivity of variational quantum
  algorithms.
\newblock {\em Phys. Rev. Lett.}, 128(8):080506, 2022.

\bibitem{du2020learnability}
Yuxuan Du, Min-Hsiu Hsieh, Tongliang Liu, Shan You, and Dacheng Tao.
\newblock Learnability of quantum neural networks.
\newblock {\em PRX Quantum}, 2(4):040337, 2021.

\bibitem{cerezo2020cost}
Marco Cerezo, Akira Sone, Tyler Volkoff, Lukasz Cincio, and Patrick~J Coles.
\newblock Cost function dependent barren plateaus in shallow parametrized
  quantum circuits.
\newblock {\em Nat Commun}, 12(1):1--12, 2021.

\bibitem{mcclean2018barren}
Jarrod~R McClean, Sergio Boixo, Vadim~N Smelyanskiy, Ryan Babbush, and Hartmut
  Neven.
\newblock Barren plateaus in quantum neural network training landscapes.
\newblock {\em Nat Commun}, 9(1):1--6, 2018.

\bibitem{sweke2019stochastic}
Ryan Sweke, Frederik Wilde, Johannes Meyer, Maria Schuld, Paul~K F{\"a}hrmann,
  Barth{\'e}l{\'e}my Meynard-Piganeau, et~al.
\newblock Stochastic gradient descent for hybrid quantum-classical
  optimization.
\newblock {\em Quantum}, 4:314, 2020.

\bibitem{wang2020noise}
Samson Wang, Enrico Fontana, Marco Cerezo, Kunal Sharma, Akira Sone, Lukasz
  Cincio, et~al.
\newblock Noise-induced barren plateaus in variational quantum algorithms.
\newblock {\em Nat Commun}, 12(1):1--11, 2021.

\bibitem{temme2017error}
Kristan Temme, Sergey Bravyi, and Jay~M Gambetta.
\newblock Error mitigation for short-depth quantum circuits.
\newblock {\em Phys. Rev. Lett.}, 119(18):180509, 2017.

\bibitem{endo2018practical}
Suguru Endo, Simon~C Benjamin, and Ying Li.
\newblock Practical quantum error mitigation for near-future applications.
\newblock {\em Phys. Rev. X}, 8(3):031027, 2018.

\bibitem{li2017efficient}
Ying Li and Simon~C Benjamin.
\newblock Efficient variational quantum simulator incorporating active error
  minimization.
\newblock {\em Phys. Rev. X}, 7(2):021050, 2017.

\bibitem{mcclean2017hybrid}
Jarrod~R McClean, Mollie~E Kimchi-Schwartz, Jonathan Carter, and Wibe~A
  De~Jong.
\newblock Hybrid quantum-classical hierarchy for mitigation of decoherence and
  determination of excited states.
\newblock {\em Phys. Rev. A}, 95(4):042308, 2017.

\bibitem{strikis2020learning}
Armands Strikis, Dayue Qin, Yanzhu Chen, Simon~C Benjamin, and Ying Li.
\newblock Learning-based quantum error mitigation.
\newblock {\em PRX Quantum}, 2(4):040330, 2021.

\bibitem{czarnik2020error}
Piotr Czarnik, Andrew Arrasmith, Patrick~J Coles, and Lukasz Cincio.
\newblock Error mitigation with clifford quantum-circuit data.
\newblock {\em Quantum}, 5:592, 2021.

\bibitem{chivilikhin2020mog}
D~Chivilikhin, A~Samarin, V~Ulyantsev, I~Iorsh, AR~Oganov, and O~Kyriienko.
\newblock Mog-vqe: Multiobjective genetic variational quantum eigensolver.
\newblock {\em arXiv preprint arXiv:2007.04424}, 2020.

\bibitem{li2020quantum}
Li~Li, Minjie Fan, Marc Coram, Patrick Riley, Stefan Leichenauer, et~al.
\newblock Quantum optimization with a novel gibbs objective function and ansatz
  architecture search.
\newblock {\em Phys. Rev. Research}, 2(2):023074, 2020.

\bibitem{ostaszewski2021structure}
Mateusz Ostaszewski, Edward Grant, and Marcello Benedetti.
\newblock Structure optimization for parameterized quantum circuits.
\newblock {\em Quantum}, 5:391, 2021.

\bibitem{grant2019initialization}
Edward Grant, Leonard Wossnig, Mateusz Ostaszewski, and Marcello Benedetti.
\newblock An initialization strategy for addressing barren plateaus in
  parametrized quantum circuits.
\newblock {\em Quantum}, 3:214, 2019.

\bibitem{skolik2021layerwise}
Andrea Skolik, Jarrod~R McClean, Masoud Mohseni, Patrick van~der Smagt, and
  Martin Leib.
\newblock Layerwise learning for quantum neural networks.
\newblock {\em Quantum Mach. Intell.}, 3(1):1--11, 2021.

\bibitem{zhang2021toward}
Kaining Zhang, Min-Hsiu Hsieh, Liu Liu, and Dacheng Tao.
\newblock Toward trainability of deep quantum neural networks.
\newblock {\em arXiv preprint arXiv:2112.15002}, 2021.

\bibitem{bittel2021training}
Lennart Bittel and Martin Kliesch.
\newblock Training variational quantum algorithms is np-hard.
\newblock {\em Phys. Rev. Lett.}, 127(12):120502, 2021.

\bibitem{elsken2019neural}
Thomas Elsken, Jan~Hendrik Metzen, and Frank Hutter.
\newblock Neural architecture search: A survey.
\newblock {\em Journal of Machine Learning Research}, 20:1--21, 2019.

\bibitem{marrero2020entanglement}
Carlos~Ortiz Marrero, M{\'a}ria Kieferov{\'a}, and Nathan Wiebe.
\newblock Entanglement-induced barren plateaus.
\newblock {\em PRX Quantum}, 2(4):040316, 2021.

\bibitem{patti2020entanglement}
Taylor~L Patti, Khadijeh Najafi, Xun Gao, and Susanne~F Yelin.
\newblock Entanglement devised barren plateau mitigation.
\newblock {\em arXiv preprint arXiv:2012.12658}, 2020.

\bibitem{Tobias2021capacity}
Tobias Haug, Kishor Bharti, and M.S. Kim.
\newblock Capacity and quantum geometry of parametrized quantum circuits.
\newblock {\em PRX Quantum}, 2:040309, Oct 2021.

\bibitem{huang2021power}
Hsin-Yuan Huang, Michael Broughton, Masoud Mohseni, Ryan Babbush, Sergio Boixo,
  Hartmut Neven, et~al.
\newblock Power of data in quantum machine learning.
\newblock {\em Nat Commun}, 12(1):1--9, 2021.

\bibitem{du2018implementable}
Yuxuan Du, Min-Hsiu Hsieh, Tongliang Liu, and Dacheng Tao.
\newblock A grover-search based quantum learning scheme for classification.
\newblock {\em New J. Phys.}, 23(2):023020, 2021.

\bibitem{cong2019quantum}
Iris Cong, Soonwon Choi, and Mikhail~D Lukin.
\newblock Quantum convolutional neural networks.
\newblock {\em Nat. Phys.}, 15(12):1273--1278, 2019.

\bibitem{wang2021towards}
Xinbiao Wang, Yuxuan Du, Yong Luo, and Dacheng Tao.
\newblock Towards understanding the power of quantum kernels in the nisq era.
\newblock {\em Quantum}, 5:531, 2021.

\bibitem{larose2019variational}
Ryan LaRose, Arkin Tikku, {\'E}tude O’Neel-Judy, Lukasz Cincio, and Patrick~J
  Coles.
\newblock Variational quantum state diagonalization.
\newblock {\em npj Quantum Inf}, 5(1):1--10, 2019.

\bibitem{du2019efficient}
Xu-Fei Yin, Yuxuan Du, Yue-Yang Fei, Rui Zhang, Li-Zheng Liu, Yingqiu Mao,  et~al.
\newblock Efficient bipartite entanglement detection scheme with a quantum
  adversarial solver.
\newblock {\em Phys. Rev. Lett.}, 128(11):110501, 2022.

\bibitem{bergholm2018pennylane}
Ville Bergholm, Josh Izaac, Maria Schuld, Christian Gogolin, Carsten Blank,
  Keri McKiernan, et~al.
\newblock Pennylane: Automatic differentiation of hybrid quantum-classical
  computations.
\newblock {\em arXiv preprint arXiv:1811.04968}, 2018.

\bibitem{Qiskit}
Qiskit: An open-source framework for quantum computing, 2019.

\bibitem{o2016scalable}
Peter~JJ O’Malley, Ryan Babbush, Ian~D Kivlichan, Jonathan Romero, Jarrod~R
  McClean, Rami Barends,  et~al.
\newblock Scalable quantum simulation of molecular energies.
\newblock {\em Phys. Rev. X}, 6(3):031007, 2016.

\bibitem{mcardle2020quantum}
Sam McArdle, Suguru Endo, Alan Aspuru-Guzik, Simon~C Benjamin, and Xiao Yuan.
\newblock Quantum computational chemistry.
\newblock {\em Rev. Mod. Phys.}, 92(1):015003, 2020.

\bibitem{yao2020reinforcement}
Jiahao Yao, Lin Lin, and Marin Bukov.
\newblock Reinforcement learning for many-body ground-state preparation
  inspired by counterdiabatic driving.
\newblock {\em Phys. Rev. X}, 11(3):031070, 2021.

\bibitem{goodfellow2016deep}
Ian Goodfellow, Yoshua Bengio, and Aaron Courville.
\newblock {\em Deep learning}.
\newblock MIT press, 2016.

\bibitem{pham2018efficient}
Hieu Pham, Melody Guan, Barret Zoph, Quoc Le, and Jeff Dean.
\newblock Efficient neural architecture search via parameters sharing.
\newblock In {\em International Conference on Machine Learning}, pages
  4095--4104, 2018.

\bibitem{huang2021greedynasv2}
Tao Huang, Shan You, Fei Wang, Chen Qian, Changshui Zhang, Xiaogang Wang, et~al.
\newblock Greedynasv2: Greedier search with a greedy path filter.
\newblock {\em arXiv preprint arXiv:2111.12609}, 2021.

\bibitem{liu2018progressive}
Chenxi Liu, Barret Zoph, Maxim Neumann, Jonathon Shlens, Wei Hua, Li-Jia Li, et~al.
\newblock Progressive neural architecture search.
\newblock In {\em Proceedings of the European Conference on Computer Vision
  (ECCV)}, pages 19--34, 2018.

\bibitem{you2020greedynas}
Shan You, Tao Huang, Mingmin Yang, Fei Wang, Chen Qian, and Changshui Zhang.
\newblock Greedynas: Towards fast one-shot nas with greedy supernet.
\newblock In {\em Proceedings of the IEEE/CVF Conference on Computer Vision and
  Pattern Recognition}, pages 1999--2008, 2020.

\bibitem{yang2020ista}
Yibo Yang, Hongyang Li, Shan You, Fei Wang, Chen Qian, and Zhouchen Lin.
\newblock Ista-nas: Efficient and consistent neural architecture search by
  sparse coding.
\newblock {\em Advances in Neural Information Processing Systems},
  33:10503--10513, 2020.

\bibitem{bubeck2012regret}
S{\'e}bastien Bubeck and Nicolo Cesa-Bianchi.
\newblock Regret analysis of stochastic and nonstochastic multi-armed bandit
  problems.
\newblock {\em Machine Learning}, 5(1):1--122, 2012.

\bibitem{gerchinovitz2016refined}
S{\'e}bastien Gerchinovitz and Tor Lattimore.
\newblock Refined lower bounds for adversarial bandits.
\newblock In {\em Advances in Neural Information Processing Systems}, pages
  1198--1206, 2016.

\bibitem{pirie2004spearman}
W~Pirie.
\newblock Spearman rank correlation coefficient.
\newblock {\em Encyclopedia of statistical sciences}, 12, 2004.

\bibitem{kendall1938measure}
M.~G. Kendall.
\newblock A new measure of rank correlation.
\newblock {\em Biometrika}, 30(1/2):81--93, 1938.

\bibitem{deb2002fast}
Kalyanmoy Deb, Amrit Pratap, Sameer Agarwal, and TAMT Meyarivan.
\newblock A fast and elitist multiobjective genetic algorithm: Nsga-ii.
\newblock {\em IEEE transactions on evolutionary computation}, 6(2):182--197,
  2002.

\bibitem{rattew2019domain}
Arthur~G Rattew, Shaohan Hu, Marco Pistoia, Richard Chen, and Steve Wood.
\newblock A domain-agnostic, noise-resistant, hardware-efficient evolutionary
  variational quantum eigensolver.
\newblock {\em arXiv preprint arXiv:1910.09694}, 2019.

\bibitem{sutton1998introduction}
Richard~S Sutton et~al.
\newblock {\em Introduction to reinforcement learning}, volume 135.
\newblock MIT Press, 1998.

\bibitem{volkoff2021large}
Tyler Volkoff and Patrick~J Coles.
\newblock Large gradients via correlation in random parameterized quantum
  circuits.
\newblock {\em Quantum Sci. Technol.}, 6(2):025008, 2021.

\bibitem{grimsley2019adaptive}
Harper~R Grimsley, Sophia~E Economou, Edwin Barnes, and Nicholas~J Mayhall.
\newblock An adaptive variational algorithm for exact molecular simulations on
  a quantum computer.
\newblock {\em Nat Commun}, 10(1):1--9, 2019.

\bibitem{tang2021qubit}
Ho~Lun Tang, VO~Shkolnikov, George~S Barron, Harper~R Grimsley, Nicholas~J
  Mayhall, Edwin Barnes, et~al.
\newblock qubit-adapt-vqe: An adaptive algorithm for constructing
  hardware-efficient ans{\"a}tze on a quantum processor.
\newblock {\em PRX Quantum}, 2(2):020310, 2021.

\bibitem{zhu2020adaptive}
Linghua Zhu, Ho~Lun Tang, George~S Barron, FA~Calderon-Vargas, Nicholas~J
  Mayhall, Edwin Barnes, et~al.
\newblock An adaptive quantum approximate optimization algorithm for solving
  combinatorial problems on a quantum computer.
\newblock {\em arXiv preprint arXiv:2005.10258}, 2020.

\bibitem{campos2021abrupt}
Ernesto Campos, Aly Nasrallah, and Jacob Biamonte.
\newblock Abrupt transitions in variational quantum circuit training.
\newblock {\em Phys. Rev. A}, 103(3):032607, 2021.

\end{thebibliography}
\end{document}